\documentstyle[eqsecnum,preprint,aps,epsf]{revtex}
\tightenlines

\newcommand{\nn}{\nonumber}
\def\lsim{\raise0.3ex\hbox{$<$\kern-0.75em\raise-1.1ex\hbox{$\sim$}}}
\def\gsim{\raise0.3ex\hbox{$>$\kern-0.75em\raise-1.1ex\hbox{$\sim$}}}

\newcommand{\ovl}[1]{\overline{#1}}

\newcommand{\eqn}[1]{(\ref{#1})}

\newcommand{\pslash}{p\kern-1ex /}
\newcommand{\Dslash}{{\cal D}\kern-1.5ex /}
\newcommand{\bpsi}{{\overline{\psi}}}
\newcommand{\bq}{{\overline{q}}}

\newcommand{\spr}{{s^\prime}}
\newcommand{\msbar}{{\overline {\rm MS}}}

\newcommand{\VEV}[3]{\left\langle #1\left| #2 \right| #3\right\rangle}

\begin{document}


\title{
\vspace{-3.0cm}
\begin{flushright}
{\normalsize hep-lat/0105020}\\
{\normalsize UTHEP-441}\\
{\normalsize UTCCP-P-103}\\
\end{flushright}
Kaon B parameter from quenched domain-wall QCD}

\author{CP-PACS Collaboration : 
  A.~Ali Khan$^{1}$\thanks{\scriptsize address till 31 August, 2000}, 
  S.~Aoki$^{2}$, 
  Y.~Aoki$^{1,2}$\thanks{\scriptsize present address:
        RIKEN BNL Research Center, Brookhaven National
        Laboratory, Upton, NY 11973, USA.},
  R.~Burkhalter$^{1,2}$, 
  S.~Ejiri$^{1}$\thanks{\scriptsize present address :
            Department of Physics, University of Wales, 
            Swansea SA2 8PP, UK.},
  M.~Fukugita$^{3}$, 
  S.~Hashimoto$^{4}$, 
  N.~Ishizuka$^{1,2}$, 
  Y.~Iwasaki$^{1,2}$, 
  T.~Izubuchi$^{5}$\thanks{\scriptsize on leave at
        Department of Physics,
	Brookhaven National Laboratory, Upton, NY 11973-5000, USA.},
  K.~Kanaya$^{2}$, 
  T.~Kaneko$^{4}$, 
  Y.~Kuramashi$^{4}$, 
  K.~I.~Nagai$^{1}$\thanks{\scriptsize present address :
            CERN, Theory Division, CH--1211 Geneva 23, Switzerland},
  M.~Okawa$^{4}$,
  H.P.~Shanahan$^{1}$\thanks{\scriptsize present address :
        Department of Biochemistry and Molecular
        Biology, University College London, London, England, UK.},
  Y.~Taniguchi$^{2}$, 
  A.~Ukawa$^{1,2}$ and 
  T.~Yoshi\'e$^{1,2}$
}
\address{
  ${}^{1}$Center for Computational Physics, University of Tsukuba,
  Tsukuba, Ibaraki 305-8577, Japan\\
  ${}^{2}$Institute of Physics, University of Tsukuba,
  Tsukuba, Ibaraki 305-8571, Japan\\
  ${}^{3}$Institute for Cosmic Ray Research, University of Tokyo,
  Kashiwa 277-8582, Japan\\
  ${}^{4}$High Energy Accelerator Research Organization (KEK),
  Tsukuba, Ibaraki 305-0801, Japan\\
  ${}^{5}$Institute of Theoretical Physics, Kanazawa University,
  Ishikawa 920-1192, Japan
}

\date{\today}

\maketitle

\begin{abstract}

We report on a calculation of $B_K$ with domain wall 
fermion action in quenched QCD. Simulations are made with a renormalization 
group improved gauge action at $\beta=2.6$ and $2.9$ corresponding to 
$a^{-1}\approx 2$GeV and $3$GeV.  Effects due to finite fifth
dimensional size $N_5$ and finite spatial size $N_\sigma$ are examined 
in detail.  Matching to the continuum operator is made perturbatively
at one loop order.  We obtain $B_K(\mu = 2 \mbox{GeV})= 0.5746(61)(191)$,
where the first error is statistical and the second error represents an 
estimate of scaling violation and ${\cal O}(\alpha^2)$ errors in the
 renormalization factor added in quadrature, 
as an estimate of the continuum value in the $\msbar$ 
scheme with naive dimensional regularization. 
This value is consistent, albeit somewhat small,  with
$B_K(\mu = 2 \mbox{GeV})= 0.628(42)$ obtained by the JLQCD Collaboration
 using the Kogut-Susskind quark action.
Results for light quark masses are also reported.

\end{abstract}



\section{Introduction}

The kaon B parameter $B_K$ is an important quantity to pin down 
the Cabibbo-Kobayashi-Maskawa matrix from experiment, 
thereby advancing our understanding of CP violation in the Standard 
Model \cite{Lellouch2000}.

A crucial ingredient in a precision calculation of $B_K$
is chiral symmetry.  Without this symmetry the relevant 
$\Delta S=2$ four-quark operator mixes with other operators of different 
chiralities.  It is a non-trivial task to accurately determine 
the mixing coefficients.  

This problem has caused significant difficulties with calculations 
using the Wilson-type fermion action, which has explicit chiral symmetry 
breaking.
While several non-perturbative methods have been developed 
to determine the mixing coefficients \cite{NPR,GBS1997,jlqcd-wilson}, 
the numerical errors in the values of $B_K$ obtained with these methods 
are still quite large \cite{Lellouch2000}.

The situation is better with the Kogut-Susskind fermion 
action for which $U(1)$ subgroup of chiral symmetry, valid at finite 
lattice spacings, ensures the correct chiral behavior of the matrix 
element\cite{STAG}.  
Exploiting this feature a systematic and extensive set of simulations 
have been carried out\cite{jlqcd}. 
Taking into account ${\cal O}(a^2)$ scaling violation and 
${\cal O}(\alpha_{\rm \ovl{MS}}^2)$ errors that arise with the use of
one-loop perturbative renormalization factors, 
$B_K(\mu=2{\rm GeV})=0.628(42)$ has been obtained in the continuum limit 
in the $\msbar$ scheme with naive dimensional regularization (NDR). 

Recent development of the domain wall\cite{Kaplan92,Shamir93,Shamir95} 
and overlap\cite{NN94,Neuberger98} fermion formalisms
has opened a prospect toward an even better calculation.
Even at finite lattice spacings, 
these formulations maintain both flavor and chiral symmetries,
either of which is broken in the Wilson-type fermion action or the
KS quark action.
Hence one expects that systematic as well as statistical uncertainties
are better controlled in these formulations than others.
A pioneering calculation of the $B_K$ parameter in this direction 
was made in Ref.~\cite{blum-soni} using the domain wall fermion formalism 
of QCD (DWQCD). 
In this article we present results of our study toward a precision 
determination of $B_K$ with DWQCD.

Our investigation is carried out in the quenched approximation 
using Shamir's formulation of domain wall fermion for quarks\cite{Shamir95}, 
and a renormalization group (RG)
improved gauge action for gluons\cite{Iwasaki83}. 
The latter choice is motivated by the result\cite{cppacs-dwf} 
that chiral symmetry is much better realized with this action than for 
the plaquette gauge action. 
We may also expect that scaling violation in $B_K$ arising from the 
gauge action is improved with the use of the RG-improved action.

We examine effects due to finite fifth dimensional size $N_5$ and 
finite spatial size $N_\sigma$ in detail.
Scaling behavior of $B_K$ is studied by adopting $\beta=2.6$
and $2.9$ corresponding to the lattice spacing $a^{-1}\approx 2$GeV and
$3$GeV. 
Matching to the continuum operator is made perturbatively at one loop
order.  
Making a constant fit in $a$ for the continuum extrapolation, we obtain 
$B_K(\mu = 2 \mbox{GeV})= 0.5746(61)(191)$
as an estimate of the continuum value in the $\msbar$ 
scheme with naive dimensional regularization (NDR).
Here the first error is statistical and the second error is an estimated
systematic error due to scaling violation and 
${\cal O}(\alpha_{\rm \ovl{MS}}^2)$ terms in the renormalization factors.
This value is consistent with the Kogut-Susskind result quoted above, 
albeit lying at the lower edge of the one standard deviation error band 
of the latter result.  
We also report on light quark masses obtained from meson mass
measurements in our simulation.   

This paper is organized as follows.
In section~\ref{sec:action} we define the fermion and gluon actions, 
where we recapitulate the argument for choosing the RG-improved action
for gluons. 
Numerical simulations and run parameters 
are described in section~\ref{sec:strategy}. 
In section~\ref{sec:renormalization} 
we discuss the operator matching between the lattice and
continuum.  Hadron mass results, in particular the chiral
behavior of pseudo scalar meson mass, are discussed in
section~\ref{sec:hadron}.  
Our main results for the kaon B parameter are given in
section~\ref{sec:BK}. 
Section~\ref{sec:ms} is devoted to the derivation of light quark mass.
We close the paper with a brief summary and comments 
in section~\ref{sec:conclusion}.

\section{Action}
\label{sec:action}

We employ Shamir's domain-wall fermion action\cite{Shamir93,Shamir95}. 
Flipping the sign of the Wilson term and the domain wall height $M$, 
we write 
\begin{eqnarray}
S_f &=& -\sum_{x,s,y,\spr} \bpsi(x,s) D_{dwf}(x,s;y,\spr) \psi(y,\spr)
 +\sum_x m_f \bq(x)  q(x)~,
\label{eq:action}
\\
D_{dwf}(x,s;y,\spr) &=& 
 D^4(x,y)\delta_{s,\spr} 
+D^5(s,\spr)\delta_{x,y}
+(M-5)\delta_{x,y} \delta_{s,\spr}~,
\label{eq:qm_dwf}
\\
D^4(x,y) &=& \sum_\mu
\frac{1}{2}\left[(1-\gamma_\mu)U_{x,\mu}\delta_{x+\hat\mu,y}
          + (1+\gamma_\mu)U^\dagger_{y,\mu}\delta_{x-\hat\mu,y}\right],
\\
D^5(s,\spr) &=& \left\{
\begin{array}{lll}
P_L\delta_{2,\spr} &\ & (s=1)\\
P_L\delta_{s+1,\spr} & +P_R\delta_{s-1,\spr} & (1<s<N_5)\\
&P_R\delta_{N_5-1,\spr} & (s=N_5)
\end{array}
\right.~,
\end{eqnarray}
where $x,y$ are four-dimensional space-time coordinates, and $s,s'$ are 
fifth-dimensional or ``flavor'' indices,  bounded as 
$1 \le s, s' \le N_5$ with the free boundary condition at both ends
(we assume $N_5$ to be even);
$P_{R/L}$ is the projection matrix $P_{R/L}=(1\pm\gamma_5)/2$, 
and $m_f$ is the bare quark mass.
The four-quark operator for our calculation is constructed with
the 4-dimensional quark field defined on the edges of the
fifth dimensional space,
\begin{eqnarray}
q(x) = P_L \psi(x,1) + P_R \psi(x,{N_5}),
\nn \\
\ovl{q}(x) = \bpsi(x,{N_5}) P_L + \bpsi(x,1) P_R.
\label{eq:quark}
\end{eqnarray}

For the gauge part of the action we employ the following form in 4
dimensions:
\begin{equation}
S_{\rm gluon} = \frac{1}{g^2}\left\{
c_0 \sum_{plaquette} {\rm Tr}U_{pl}
+ c_1  \sum_{rectangle} {\rm Tr} U_{rtg}
+ c_2 \sum_{chair} {\rm Tr} U_{chr}
+ c_3 \sum_{parallelogram} {\rm Tr} U_{plg}\right\}, 
\label{eqn:RG}
\end{equation}
where the first term represents the standard plaquette action, and the 
remaining terms are six-link loops formed by a $1\times 2$ rectangle, 
a bent $1\times 2$ rectangle (chair) and a 3-dimensional parallelogram. 
The coefficients $c_0, \cdots, c_3$ satisfy the normalization condition
\begin{equation}
c_0+8c_1+16c_2+8c_3=1. 
\end{equation}
The RG-improved action of Iwasaki\cite{Iwasaki83} is defined by
setting the parameters to $c_0=3.648, c_1=-0.331, c_2=c_3=0$.
With this choice of parameters the action is expected to 
exhibit smooth gauge field fluctuations approximating those in the
continuum limit better than with the unimproved plaquette action. 

A basic piece of information for our study of $B_K$ with DWQCD 
is in what range of the coupling constant $\beta=6/g^2$ and domain wall 
height $M$ DWQCD realizes exact chiral symmetry in the limit of infinite 
fifth dimensional size $N_5\to\infty$.  
This point has been examined in a number of recent studies 
\cite{cppacs-dwf,RBC0007038}. 
Investigations using the axial vector Ward-Takahashi identity show 
that a non-zero residual quark mass $m_{5q}$, 
which represents chiral symmetry breaking, remains
even in the limit of infinite fifth dimensional size $N_5\to\infty$ 
if the lattice spacing is as coarse as $a^{-1}\approx $1~GeV. 

The chiral property is much improved as the coupling constant is 
decreased.  In the range corresponding to $a^{-1}\sim2$ GeV, 
the value of residual quark mass becomes an order of magnitude smaller 
than at $a^{-1}\sim1$ GeV at similar fifth dimensional sizes $N_5$. 
For the standard plaquette gauge action, it is still not clear 
whether $m_{5q}$ vanishes exponentially with a small decay 
rate \cite{RBC0007038} or remains finite, albeit very small, 
as $N_5\to\infty$ \cite{cppacs-dwf}. 
In contrast, for the RG-improved gauge action, the residual quark mass 
shows an $N_5$ dependence consistent with an exponential decay in $N_5$ 
up to $N_5=24$.  Furthermore the magnitude of $m_{5q}$ is an order of 
magnitude smaller than that for the plaquette gauge action. 

We can conclude that chiral symmetry is much better realized with the 
RG-improved gauge action than with the plaquette gauge action.
We therefore employ the RG-improved gauge action for our investigation of 
the $B_K$ parameter.

\section{Run parameters and measurements}
\label{sec:strategy}

Parameters of our simulations and the number of configurations employed 
are summarized in Table~\ref{tab:config}.  
We carry out runs at two values of coupling,  
$\beta=2.6$ and $2.9$, corresponding 
to a lattice spacing $a^{-1}=1.81(4)$ GeV and $2.81(6)$ GeV
determined from the the $\rho$ meson mass $m_\rho=770$MeV.  
The first value is chosen since chiral symmetry is sufficiently well 
realized \cite{cppacs-dwf}, and the second value is selected to check 
scaling violation effects. 

For our main runs we use the lattice size 
$N_\sigma^3 \times N_t \times N_5 = 24^3\times 40\times 16$ 
at $\beta=2.6$, and $32^3\times 60\times 16$ at $\beta=2.9$. 
These lattices have a reasonably large spatial size of 
$aN_\sigma\approx 2.6$~fm or 2.3~fm respectively.   
The choice of $N_5=16$ at $\beta=2.6$ is based on our previous 
result\cite{cppacs-dwf} that the anomalous quark mass is 
already quite small, $m_{5q}=0.274(42)$~MeV, for this parameter set with the 
domain wall height $M=1.8$. 
In this paper the domain wall height is also taken to be $M=1.8$.

We examine the dependence on the fifth dimensional length $N_5$ 
at $\beta=2.6$ for the spatial size $N_\sigma=24$ using $N_5=16$ and 
$N_5=32$.  Since we expect the decay rate in $N_5$ to become 
larger toward weaker coupling, we only employ $N_5=16$ at $\beta=2.9$. 

The spatial size dependence is examined at $\beta=2.6$
varying the spatial size from $N_\sigma=24$ to either $N_\sigma=16$ or 32,
which correspond to the physical size of $aN_\sigma\sim1.7$ and $3.4$ fm.
The size dependence is also checked at $\beta=2.9$ by adopting $N_\sigma=24$
and $32$ ($aN_\sigma=1.7, 2.3$ fm).

We take degenerate quarks in our calculations.  
The common value of bare quark mass is chosen to be 
$m_fa=0.01, 0.02, 0.03, 0.04$ at both $\beta=2.6$ and 2.9, 
which covers the range $m_{PS}/m_V\approx 0.4 - 0.8$.

Quenched gauge configurations are generated on four-dimensional
lattices.
A sweep of gauge update contains one pseudo-heatbath and four 
overrelaxation steps.
After a thermalization of 2000 sweeps hadron propagators and 3-point 
functions necessary to evaluate $B_K$ are calculated
at every 200th sweep.
The gauge configuration on each fifth dimensional coordinate $s$
is identical and is fixed to the Coulomb gauge.

In the course of our simulation we measure the kaon B parameter,  
\begin{eqnarray}
B_K =
\frac{
\VEV{K}{\ovl{s}\gamma_\mu(1-\gamma_5)d \ovl{s}\gamma_\mu(1-\gamma_5)d}{K}}
{\frac{8}{3}\VEV{K}{\ovl{s}\gamma_\mu\gamma_5d}{0}
 \VEV{0}{\ovl{s}\gamma_\mu\gamma_5d}{K}}
\label{eqn:BK}
\end{eqnarray}
and the matrix element divided by the pseudo scalar density, 
\begin{eqnarray}
B_P =
\frac{
\VEV{K}{\ovl{s}\gamma_\mu(1-\gamma_5)d \ovl{s}\gamma_\mu(1-\gamma_5)d}{K}}
{\VEV{K}{\ovl{s}\gamma_5d}{0} \VEV{0}{\ovl{s}\gamma_5d}{K}}
\label{eqn:BP}
\end{eqnarray}
which should vanish at $m_\pi\to 0$.  
The $s$ and $d$ quark fields defining these quantities are the boundary 
fields given by (\ref{eq:quark}), and the four-quark and bilinear operators 
are taken to be local in the 4-dimensional space-time.  

The domain-wall quark propagator needed to extract the $B$ parameters 
above is calculated with the conjugate gradient algorithm with an even-odd 
pre-conditioning.  
Two quark propagators are evaluated for each configuration corresponding 
to the wall source placed at either $t=1$ or $40$ at $\beta=2.6$
($t=4$ or $57$ at $\beta=2.9$)
in the time direction with the Dirichlet boundary condition, while 
the periodic boundary condition is imposed in the spatial directions.  
The two quark propagators are combined to form the kaon Green function 
with an insertion of the four-quark operator at time slices 
$1\leq t\leq N_t$ in a standard manner
(see, {\it e.g.,} Ref.~\cite{jlqcd}). 

We employ the quark propagators above to also evaluate pseudo scalar and 
vector meson propagators, and extract their masses. 
These masses are calculated for degenerate quark-antiquark pair.  
The physical point for light quark masses $m_{ud}$ and $m_s$  
is calculated by linearly 
fitting the meson masses $m_{PS}^2$ and $m_V$ as a function of $m_f$,
and using the experimental values of $m_\pi/m_\rho$ and $m_K/m_\rho$ or
$m_\phi/m_\rho$ as input.

\section{Operator matching}
\label{sec:renormalization} 

We carry out matching of the lattice and continuum operators 
at a scale $q^*=1/a$ using one-loop perturbation theory
\cite{zfactor} and the $\msbar$ scheme with NDR in the
continuum.
The continuum value at a physical scale {\it e.g.,} $\mu=2$ GeV, is 
obtained {\it via} a renormalization group running 
from $q^*=1/a$ to $\mu$.
\begin{eqnarray}
B_K({\rm NDR}, \mu ) & = &
\left[
1- \frac{\alpha_{\overline{\rm MS}}(\mu)}{4\pi}
\frac{\gamma_1\beta_0-\gamma_0\beta_1}{2\beta_0^2}
\right]^{-1}
\left[
1- \frac{\alpha_{\overline{\rm MS}}(q^*)}{4\pi}
\frac{\gamma_1\beta_0-\gamma_0\beta_1}{2\beta_0^2}
\right] \nonumber \\
&\times& \left[\frac{\alpha_{\overline{\rm MS}}(q^*)}
 {\alpha_{\overline{\rm MS}}(\mu)}
\right]^{-\gamma_0/2\beta_0}
B_K({\rm NDR},q^*),
\label{eq:run_bk}
\end{eqnarray}
where $\beta_0=11$, $\beta_1 =102$, $\gamma_0 =4$ and $\gamma_1 = -7$
\cite{BJW90} are the $N_f=0$
quenched values for the renormalization group coefficients.

In the domain wall formalism the renormalization factor of an $n$-quark 
operator $O_n$ has a generic form
\begin{eqnarray}
&&
O_n^{\ovl{\rm MS}}(\mu) = Z O_n^{\rm lattice}(1/a),
\\&&
Z=(1-w_0^2)^{-n/2}Z_w^{-n/2}Z_{O_n},
\end{eqnarray}
where $w_0=1-M$, and $Z_w$ represents the quantum correction to the 
normalization factor $1-w_0^2$ of physical quark fields $q, \bq$, 
and $Z_{O_n}$ is the vertex correction to $O_n$.
In the present paper we need the factors 
$Z_2$, $Z_m$, $Z_A$, $Z_P$ and $Z_{O_4}$
for the quark wave function, quark mass, axial vector current, pseudo
scalar density and the four-quark $\Delta S=2$ weak operator.
Perturbative calculation of these renormalization factors at one loop 
order is given in Ref.~\cite{zfactor} for the DWQCD system with the 
standard plaquette gauge action.
Here we summarize results for the RG-improved gauge action.  

The generic form of the one-loop renormalization factors is given by
\begin{eqnarray}
&&
Z_w(\mu a) = 1+\frac{2w_0}{1-w_0^2} \frac{g^2 C_F}{16\pi^2} z_w(M),
\\&&
Z_2(\mu a)
 = 1+\frac{g^2 C_F}{16\pi^2}\left[-\log(\mu a)^2 + z_2(M)\right],
\\&&
Z_m(\mu a)
 = 1+\frac{g^2 C_F}{16\pi^2}\left[-3\log(\mu a)^2 + z_m(M)\right], 
\\&&
Z_A(\mu a) = 1+\frac{g^2 C_F}{16\pi^2}z_A(M),
\\&&
Z_P(\mu a) = 1+\frac{g^2 C_F}{16\pi^2}\left[3\log(\mu a)^2 + z_P(M)\right],
\\&&
Z_{O_4}(\mu a)=1+\frac{g^2}{16\pi^2}\left[-2\log(\mu a)^2+z_{O_4}(M)\right],
\end{eqnarray}
where $C_F$ is the second Casimir invariant $C_F=4/3$ and the finite part
$z_{O_n}$ is a function of the domain-wall height $M$.
The difference between the plaquette and the RG action resides in
the finite part.

In the first row of Table \ref{tab:zfactor} we list the finite parts of the
renormalization factors at $M=1.8$. 
The one-loop correction in $Z_w$ is very large for our choice of $M$ 
because of the tadpole factor in $z_w$ and division with 
$1-w_0^2$\cite{zfactor}.
Hence we apply a tadpole improvement by explicitly moving the one-loop
correction to the domain wall height $M$ from $Z_w$ to $w_0$ additively,
and by factoring out a tadpole factor $u^{n/2}=P^{n/8}$ with $P$ the
plaquette from $Z_{O_n}$. This leads to the rewriting, 
\begin{eqnarray}
Z \to Z^{\rm MF} =
\left(1-\left(w_0^{\rm MF}\right)^2\right)^{-n/2}
\left(Z_w^{\rm MF}\right)^{-n/2}
 u^{n/2} Z_{O_n}^{\rm MF},
\end{eqnarray}
where
\begin{eqnarray}
&&
w_0^{\rm MF} = w_0+4(1-u),
\label{eqn:mfM}
\\&&
Z_w^{\rm MF}= Z_w|_{w_0=w_0^{\rm MF}}
+\frac{4w_0^{\rm MF}}{1-(w_0^{\rm MF})^2}g^2C_Fu_1,
\\&&
Z_2^{\rm MF} =
Z_2|_{w_0=w_0^{\rm MF}}+\frac{1}{2}g^2C_Fu_1,
\\&&
Z_m^{\rm MF} =
Z_m|_{w_0=w_0^{\rm MF}}-\frac{1}{2}g^2C_Fu_1,
\\&&
Z_{O_n}^{\rm MF} =
Z_{O_n}|_{w_0=w_0^{\rm MF}}+\frac{n}{4}g^2C_Fu_1.
\end{eqnarray}
Here $u_1$ is the one-loop correction to the tadpole factor
$u=1-g^2C_Fu_1/2+\cdots$ which has the values
\begin{eqnarray}
u_1=\cases{
0.125000 & ({\rm plaquette action})\cr
0.052567 & ({\rm RG improved action})}.
\end{eqnarray}

For the tadpole factor $u = P^{1/4}$
we use the following value of the plaquette for the RG action 
\begin{eqnarray}
P=\cases{
0.670632(10) \quad{\rm at}\ \beta=2.6\cr
0.707662(5) \quad{\rm at}\ \beta=2.9},
\end{eqnarray}
obtained from our main simulations.
The domain-wall height is shifted according to \eqn{eqn:mfM} as
\begin{eqnarray}
M=1.8\to M^{\rm MF}=\cases{
1.4198 \quad{\rm for}\ \beta=2.6\cr
1.4687 \quad{\rm for}\ \beta=2.9}.
\end{eqnarray}
In the second and third rows of Table \ref{tab:zfactor} 
we list the finite parts of the renormalization factors after 
tadpole improvement.

A mean-field estimate appropriate for the RG-improved action is used for
calculating the coupling constant $g^2_{\overline{\rm MS}}(\mu)$, 
which is given with the following formula for the quenched 
case \cite{cppacs-full}
\begin{eqnarray}
\frac{1}{g^2_{\ovl{\rm MS}}(\mu)} &=&
\left(3.648 P - 2.648 R\right)\frac{\beta}{6}
+\frac{22}{16\pi^2}\log(\mu a) - 0.1006,
\label{eqn:coupling-PR}
\end{eqnarray}
where $R$ is a $1\times2$ rectangular Wilson loop whose value is given as
\begin{eqnarray}
R=\cases{
0.45283(2) \quad{\rm at}\ \beta=2.6\cr
0.50654(1) \quad{\rm at}\ \beta=2.9}.
\end{eqnarray}
The gauge coupling at $\mu=1/a$ turns out to be
\begin{eqnarray}
g^2_{\ovl{\rm MS}}(1/a)=\cases{
2.2731 \quad{\rm at}\ \beta=2.6\cr
2.0046 \quad{\rm at}\ \beta=2.9}.
\end{eqnarray}

For $B_K$ the factor $(1-w_0^2)^2Z_w^2$ cancels out, and 
the one-loop value is given by the ratio
\begin{equation}
Z_{B_K}(\mu a)=\frac{Z_{O_4}}{Z_A^2}
=\frac{1+(-2\log (\mu a)^2+z_{O_4})g^2/(16\pi^2)}{(1+(C_F z_A)g^2/(16\pi^2))^2}
=1+\frac{g^2}{16\pi^2}\left(-2\log (\mu a)^2+z_{B_K}\right).
\end{equation}
In Table \ref{tab:zbk} we give the finite parts of $Z_{B_K}$ with and
without mean field approximation at $M=1.8$ together with those for
$Z_{B_P}$,
\begin{eqnarray}
Z_{B_P}(\mu a)=1+\frac{g^2}{16\pi^2}\left(-10\log (\mu a)^2+z_{B_P}\right).
\label{eqn:ZBP}
\end{eqnarray}

The finite parts $z_{O_4}$ and $2C_Fz_A$ are very similar in magnitude, 
albeit individually not very small.  As a result the finite part 
$z_{B_K}=z_{O_4}-2C_Fz_A$ for $B_K$ is small, and 
the renormalization factor for $B_K$ with the tadpole improvement
turned out to be very near unity {\it e.g.} 
at the matching scale $q^*=1/a$,
\begin{eqnarray}
Z_{B_K}^{\ovl{\rm MS}}(q^*=1/a)=\cases{
0.984 \quad{\rm at}\ \beta=2.6\cr
0.988 \quad{\rm at}\ \beta=2.9}.
\end{eqnarray}
The $Z$ factor at the scale $\mu=2$GeV obtained with a 2-loop 
running with Eq.(\ref{eq:run_bk}) \cite{BJW90} becomes
\begin{eqnarray}
Z_{B_K}^{\ovl{\rm MS}}(\mu=2{\rm GeV})=\cases{
0.979 \quad{\rm at}\ \beta=2.6\cr
1.006 \quad{\rm at}\ \beta=2.9}.
\end{eqnarray}
Meanwhile $Z_{B_P}$ is evaluated by setting $\mu=2$ GeV in
Eq.~\eqn{eqn:ZBP}
\begin{eqnarray}
Z_{B_P}^{\ovl{\rm MS}}(\mu=2{\rm GeV})=\cases{
1.007 \quad{\rm at}\ \beta=2.6\cr
1.129 \quad{\rm at}\ \beta=2.9}.
\end{eqnarray}

For quark mass the renormalization factor at the matching scale $q^*=1/a$ 
takes the values 
\begin{eqnarray}
Z_q^{\ovl{\rm MS}}(q^*=1/a)=
\left(1-\left(w_0^{\rm MF}\right)^2\right)\left(Z_w^{\rm MF}\right)
 u^{-1} Z_m^{\rm MF}(q^*=1/a)=\cases{
1.173371\quad{\rm at}\ \beta=2.6\cr
1.094189\quad{\rm at}\ \beta=2.9}.
\end{eqnarray}
With a renormalization group running from the scale $q^*$ to $\mu=2$~GeV using 
the four-loop anomalous dimension and beta function \cite{Chetyrkin}, 
we have  
\begin{eqnarray}
m(\mu)=\frac{c(\alpha_{\ovl{\rm MS}}(\mu)/\pi)}
 {c(\alpha_{\ovl{\rm MS}}(q^*)/\pi)} m(q^*),
\end{eqnarray}
and the renormalization factor becomes
\begin{eqnarray}
Z_q^{\ovl{\rm MS}}(\mu=2 {\rm GeV})=
\cases{
1.155769\quad{\rm at}\ \beta=2.6\cr
1.147224\quad{\rm at}\ \beta=2.9}.
\end{eqnarray}
The four-loop running factor $c(\alpha_{\ovl{\rm MS}}(\mu)/\pi)$ is given
by \cite{Chetyrkin}
\begin{eqnarray}
c(x) &=& (x)^{\bar{\gamma_0}} \Biggl\{
 1 + (\bar{\gamma_1} - \bar{\beta_1}\bar{\gamma_0})x
\nn\\&+&
\frac{1}{2}
\left[
(\bar{\gamma_1} - \bar{\beta_1}\bar{\gamma_0})^2 
+
\bar{\gamma_2} + \bar{\beta_1}^2\bar{\gamma_0}
- \bar{\beta_1}\bar{\gamma_1} -\bar{\beta_2}\bar{\gamma_0}
\right] x^2
\nn\\&+&
\Biggl[ 
\frac{1}{6}(\bar{\gamma_1} - \bar{\beta_1}\bar{\gamma_0})^3
+\frac{1}{2}(\bar{\gamma_1} - \bar{\beta_1}\bar{\gamma_0})
(\bar{\gamma_2} + \bar{\beta_1}^2\bar{\gamma_0}
- \bar{\beta_1}\bar{\gamma_1} -\bar{\beta_2}\bar{\gamma_0})
\nn\\&&
+\frac{1}{3}\left(
\bar{\gamma_3}
-\bar{\beta_1}^3\bar{\gamma_0} + 2\bar{\beta_1} \bar{\beta_2}\bar{\gamma_0}
-\bar{\beta_3}\bar{\gamma_0} + \bar{\beta_1}\bar{\gamma_1}
-\bar{\beta_2}\bar{\gamma_1} - \bar{\beta_1}\bar{\gamma_2}
\right)
\Biggr] x^3 + {\cal O}(x^4)
\Biggr\},
\end{eqnarray}
where
\begin{eqnarray}
&&
\bar{\gamma_i} = \frac{\gamma_i^m}{4^i\beta_0},\quad
\bar{\beta_i} = \frac{\beta_i}{4^i\beta_0},
\\&&
\beta_0=11,\quad
\beta_1=102,\quad
\beta_2=\frac{2857}{2},\quad
\beta_3=\frac{149753}{6}+3564\zeta(3),
\\&&
\gamma_0^m = 4,\quad
\gamma_1^m = \frac{202}{3},\quad
\gamma_2^m = 1249,
\\&&
\gamma_3^m = \frac{4603055}{162}+\frac{135680}{27}\zeta(3)-8800\zeta(5)
\end{eqnarray}
with $\zeta$ the Riemann zeta-function.

Let us add a comment on the systematic error due to operator matching.
Since we have used the one-loop renormalization factor for
operator matching, the systematic error should include contributions from
higher loop corrections.
We estimate the magnitude of these corrections by 
changing the matching scale
from $q^*=1/a$ to $q^*=\pi/a$ and also 
adopting a different definition for gauge coupling using
the plaquette value only\cite{cppacs-full} given by
\begin{eqnarray}
\frac{1}{g^2_{\ovl{\rm MS}}(\mu)}=P\frac{\beta}{6}
+\frac{22}{16\pi^2}\log(\mu a)+0.2402.
\label{eqn:coupling-P}
\end{eqnarray}
The gauge coupling at $\mu=1/a$ becomes
\begin{eqnarray}
g^2_{\ovl{\rm MS}}(1/a)=\cases{
1.8839 \quad{\rm at}\ \beta=2.6\cr
1.7176 \quad{\rm at}\ \beta=2.9}.
\end{eqnarray}

\section{Pseudo scalar and vector meson masses}
\label{sec:hadron}

\subsection{Extraction of meson masses}

\begin{figure}[p]
 \begin{center}
  \leavevmode
  \epsfxsize=8cm \epsfbox{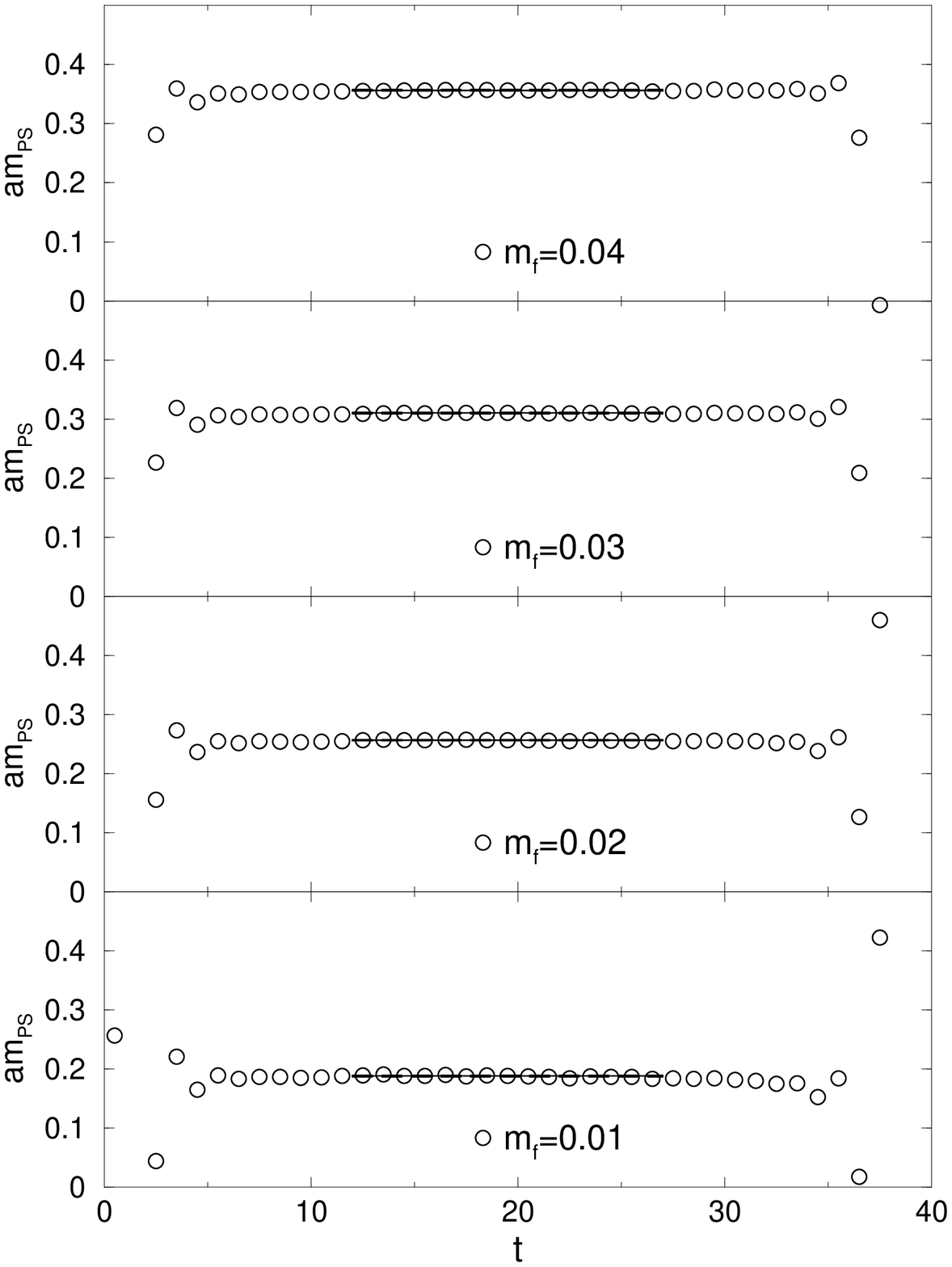}
  \epsfxsize=8cm \epsfbox{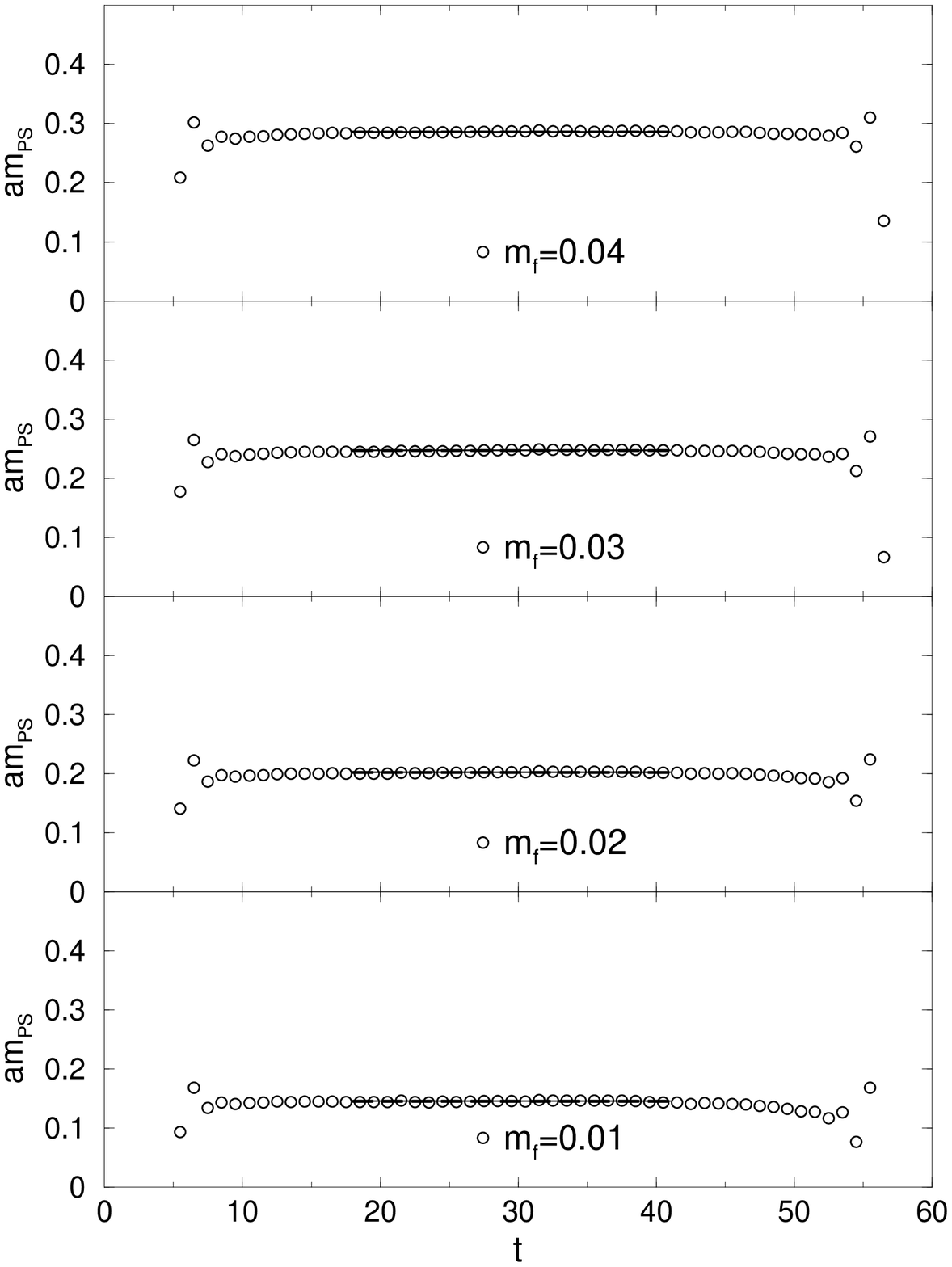}
  \caption{Effective pseudo scalar meson mass as a function of temporal
  distance $t$ at $\beta=2.6$ on a $24^3\times40\times16$ lattice (left)
  and at $\beta=2.9$ on a $32^3\times60\times16$ lattice (right).
  Lines show constant fit over the fitted range.}
  \label{fig:mpi-t}
 \end{center}
\end{figure}

\begin{figure}[p]
 \begin{center}
  \leavevmode
  \epsfxsize=8cm \epsfbox{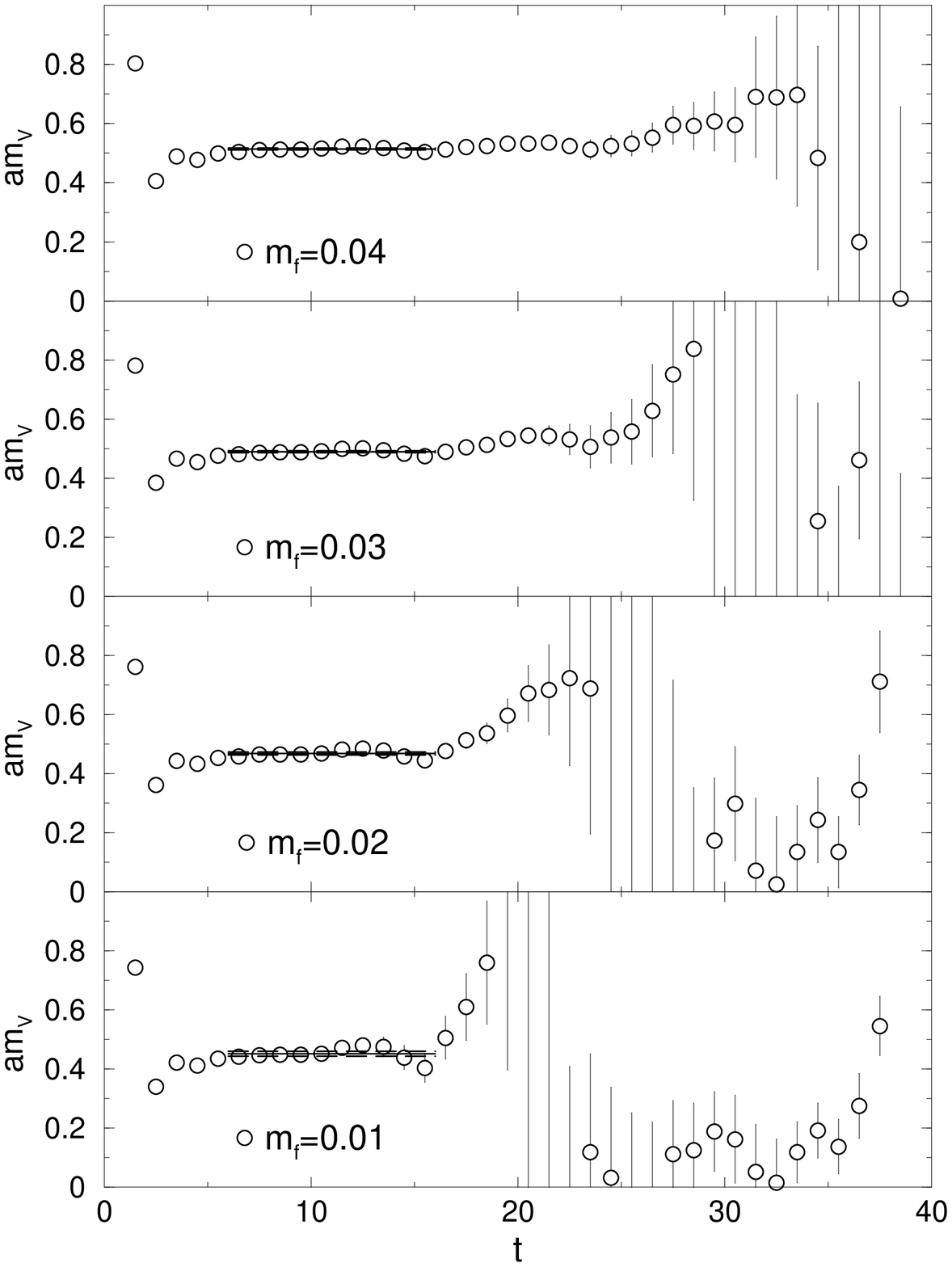}
  \epsfxsize=8cm \epsfbox{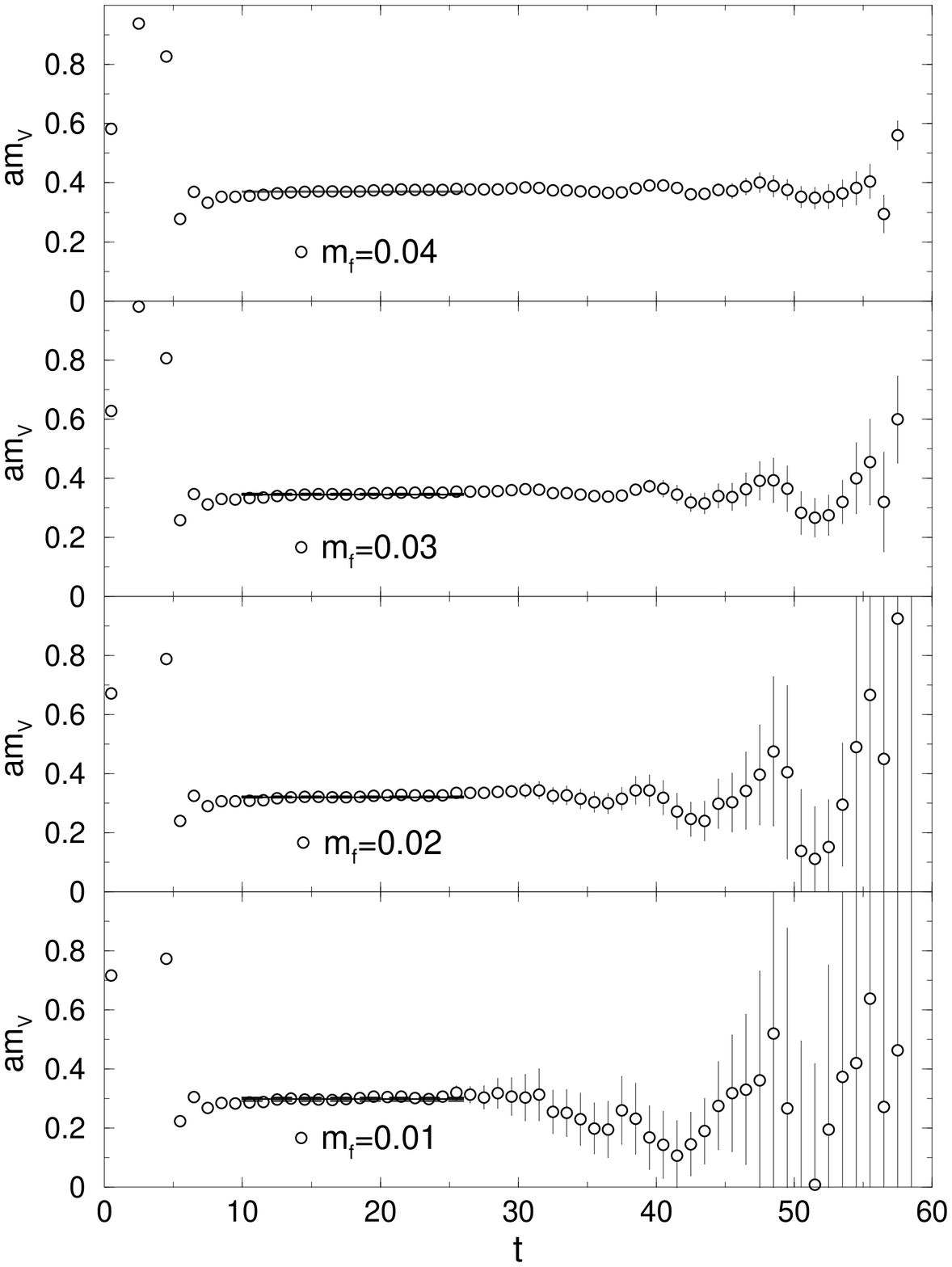}
  \caption{Effective vector meson mass as a function of temporal
  distance $t$ at $\beta=2.6$ on a $24^3\times40\times16$ lattice (left)
  and at $\beta=2.9$ on a $32^3\times60\times16$ lattice (right).
  Lines show constant fit over the fitted range.}
  \label{fig:mrho-t}
 \end{center}
\end{figure}

We extract pseudo scalar and vector meson masses $m_{PS}$ and $m_V$ 
at each $m_f$, $N_\sigma$ and $N_5$ by a single exponential
fit of meson propagators.   Representative plots of effective mass 
are shown in Figs.~\ref{fig:mpi-t} and \ref{fig:mrho-t}. 
The fitting range chosen from inspection of such plots is 
$12\le t \le27$  and $6\le t \le16$ for pseudo scalar and vector 
meson mass for all simulations at $\beta=2.6$, 
and $18\le t \le41$ and $10\le t \le 26$ at $\beta=2.9$.
In Tables \ref{tab:R26.16x40x16}-\ref{tab:R29.32x60x16} we list the
numerical values of $m_{PS} a$,
$m_V a$ and the ratio at four quark masses
$m_f a=0.01, 0.02, 0.03, 0.04$ for each set of run parameter.
The errors given 
are calculated by a single elimination jackknife procedure.

\subsection{Chiral extrapolation}

\begin{figure}[p]
 \begin{center}
  \leavevmode
  \epsfxsize=8cm \epsfbox{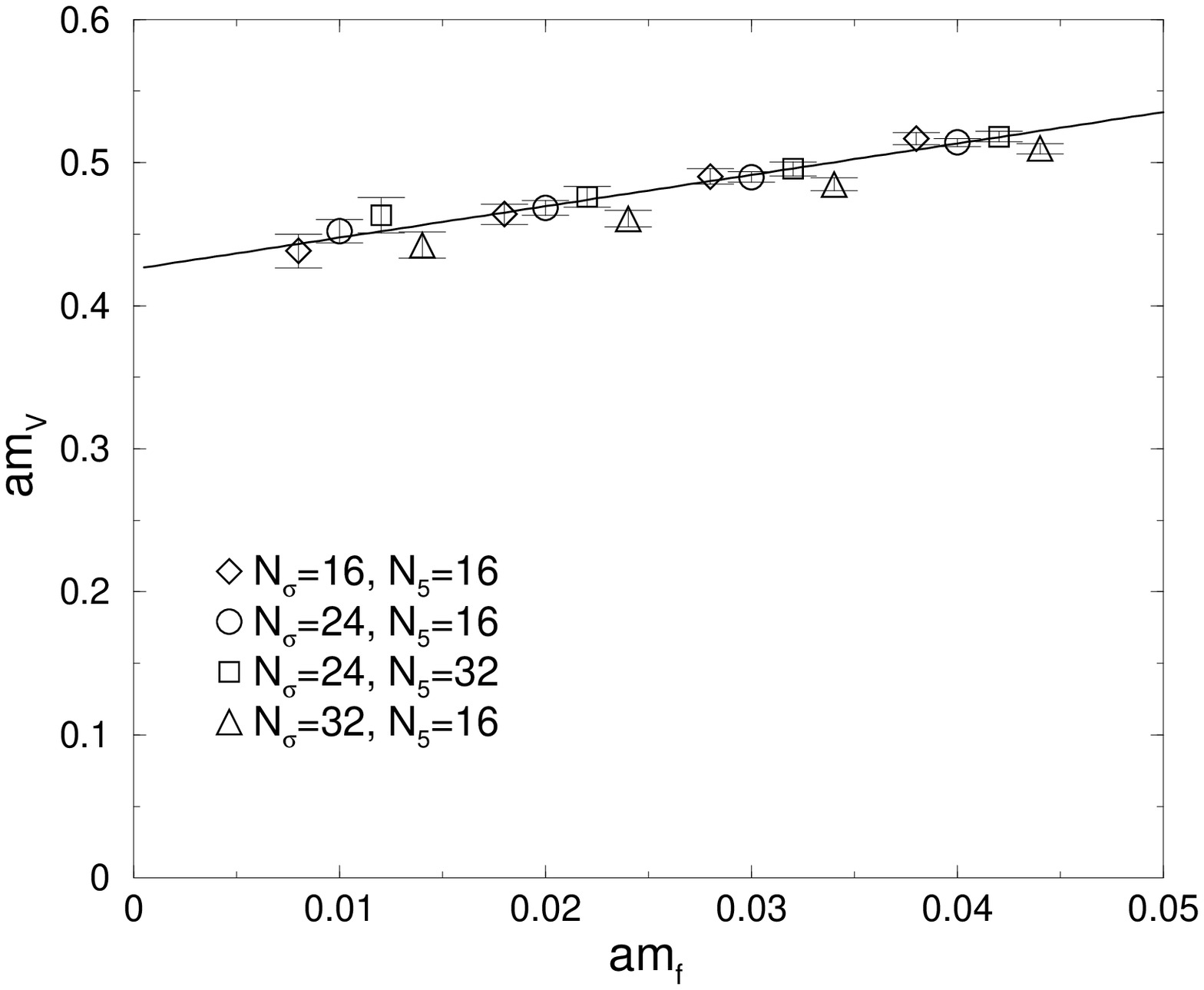}
  \epsfxsize=8cm \epsfbox{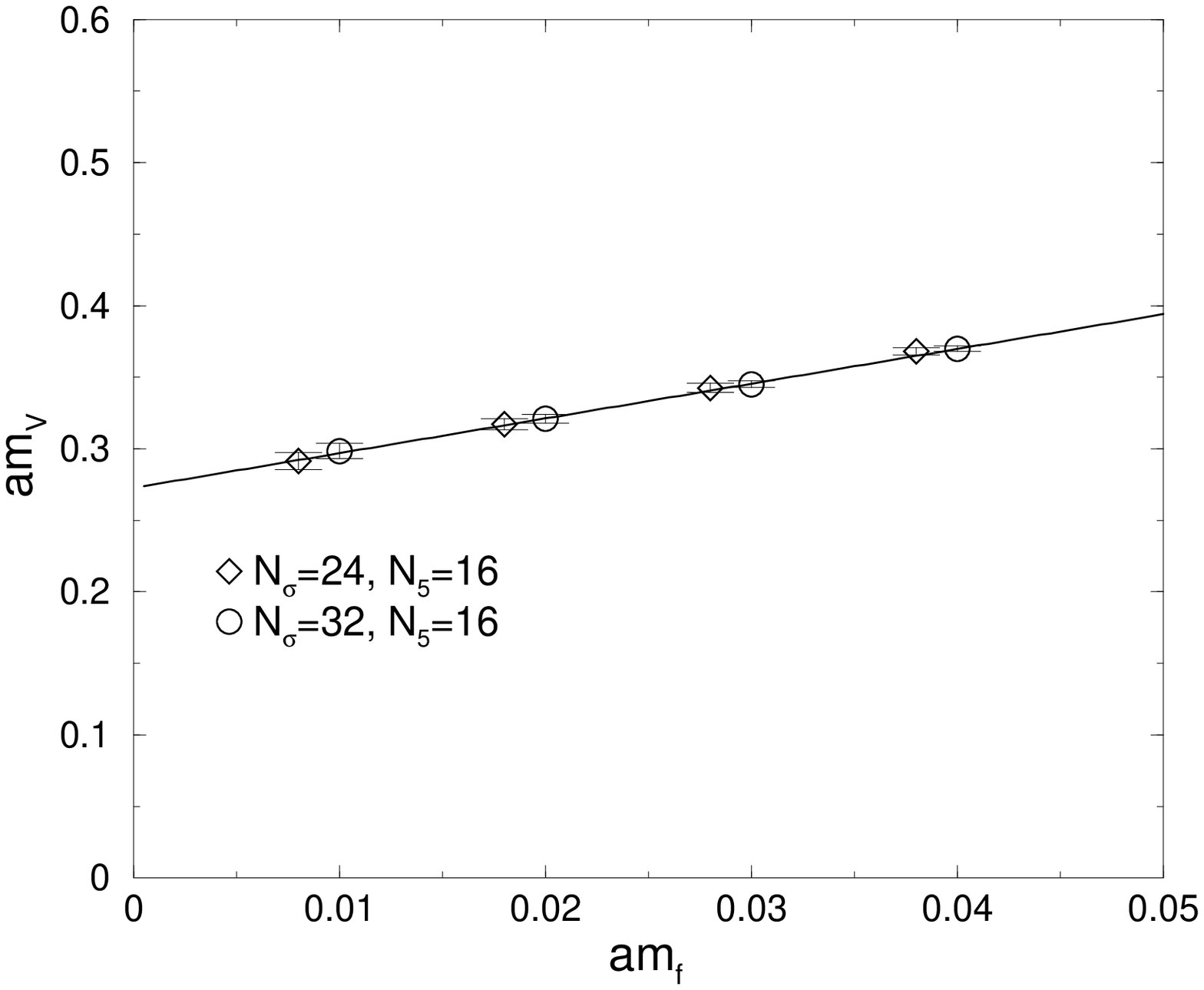}
  \caption{Vector meson mass as a function of bare quark mass $m_f$
  at $\beta=2.6$ (left) and at $\beta=2.9$ (right).
  Lines show linear fits to main runs.
  The data except for the main run are shifted in $m_f$.}
  \label{fig:mrho-mf}
 \end{center}
\end{figure}

For chiral extrapolation we fit the light hadron masses $m_{PS}^2$ 
and $m_V$ linearly as a function of $m_f a$ as illustrated in 
Figs.~\ref{fig:mpi2-mf} and \ref{fig:mrho-mf}.
Since pseudo scalar meson mass thus extrapolated does not vanish at
$m_f=0$, we employ a fit of the form 
\begin{eqnarray}
m_{PS}^2a^2 &=& A_{PS}(m_fa+m_{res}a),\label{eq:pion}\\
m_Va&=&A_V+B_Vm_fa \label{eq:rho}
\end{eqnarray}
and determine the parameters $A_{PS}, m_{res}a$ for the pseudo scalar meson, 
and $A_V, B_V$ for the vector meson.  
The physical point for the bare quark mass parameter $m_f$ corresponding 
to physical $u$ and $d$ quark ($m_f^{ud}$), 
which are assumed degenerate, and $s$ quark  ($m_f^{s}$) 
are fixed by the equations
\begin{eqnarray}
\frac{\sqrt{ A_{PS}(m_f^{ud}a+m_{res}a)}}{A_V+B_Vm_f^{ud}a}
&=&\frac{m_\pi}{m_\rho}=\frac{0.135}{0.77},\label{eq:udquark}
\\
\frac{\sqrt{ A_{PS}(m_f^{s}a/2+m_{res}a)}}{A_V+B_Vm_f^{ud}a}
&=&\frac{m_K}{m_\rho}=\frac{0.495}{0.77}, \label{eq:squark}
\\
\frac{A_V+B_Vm_f^s(\phi)a}{A_V+B_Vm_f^{ud}a}
&=&\frac{m_\phi}{m_\rho}=\frac{1.0194}{0.77},
\label{eq:spquark}
\end{eqnarray}
where for $s$ quark we employ the kaon ($m_f^s$) or phi ($m_f^s(\phi)$) meson
mass as input.
We then fix the lattice spacing $a$ by setting the vector meson mass 
at the physical quark mass point $m_f^{ud}$ to the experimental value 
$m_\rho=770$ MeV.  Numerical values of lattice spacing and other parameters 
are listed in Table \ref{tab:massresults}. 

\begin{figure}[p]
 \begin{center}
  \leavevmode
  \epsfxsize=8cm \epsfbox{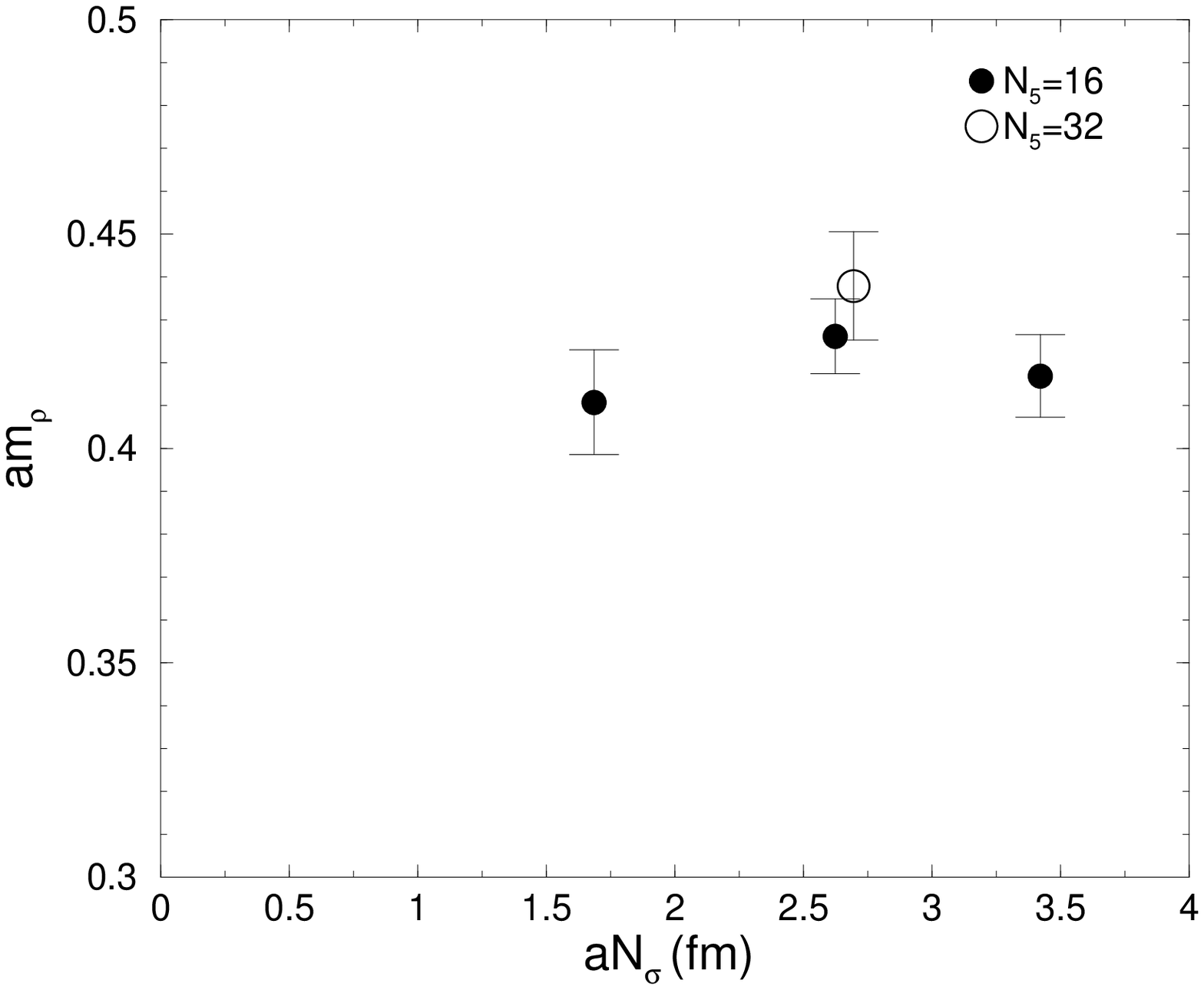}
  \epsfxsize=8cm \epsfbox{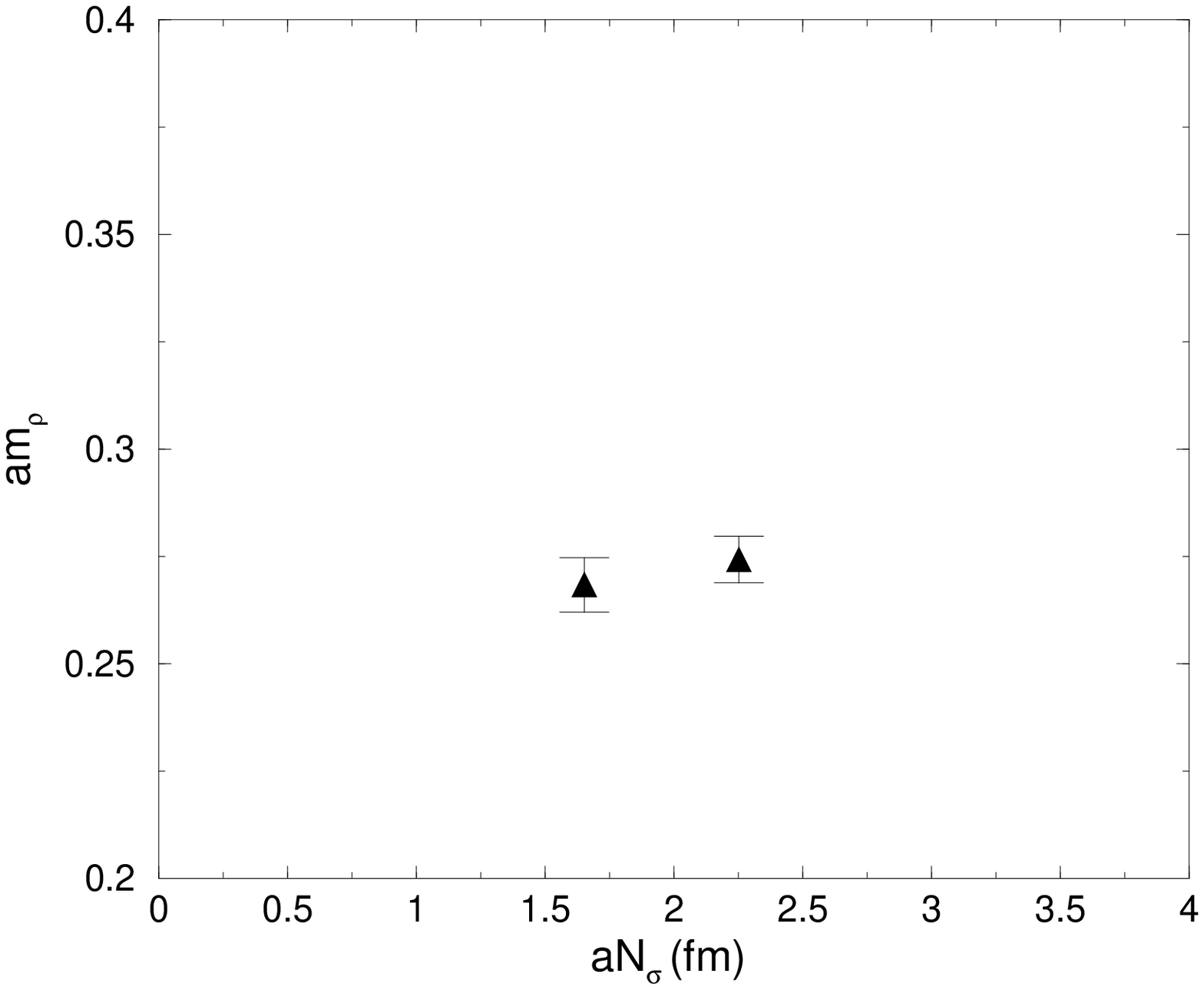}
  \caption{Rho meson mass as a function of spatial lattice size
  $N_\sigma$ at $\beta=2.6$ (left) and at $\beta=2.9$ (right).
  Filled symbols represent data at fifth dimensional length $N_5=16$
  and an open circle represents that at $N_5=32$.}
  \label{fig:mrho0-V}
 \end{center}
\end{figure}

In Fig.~\ref{fig:mrho0-V} we plot results for $m_\rho a$
at the physical point. 
The values for given $\beta$ are reasonably consistent with 
each other; 
the variation of results depending on spatial volume is mild, 
and the difference between the fifth dimensional size $N_5=16$ and $32$ 
at $\beta=2.6$ on $24^3\times40$ lattice is a one-standard deviation effect. 
In the following analyses we use the lattice spacing corresponding to each
spatial size and fifth dimensional length.  

\subsection{Chiral property of pseudo scalar meson mass}

\begin{figure}[p]
 \begin{center}
  \leavevmode
  \epsfxsize=8cm \epsfbox{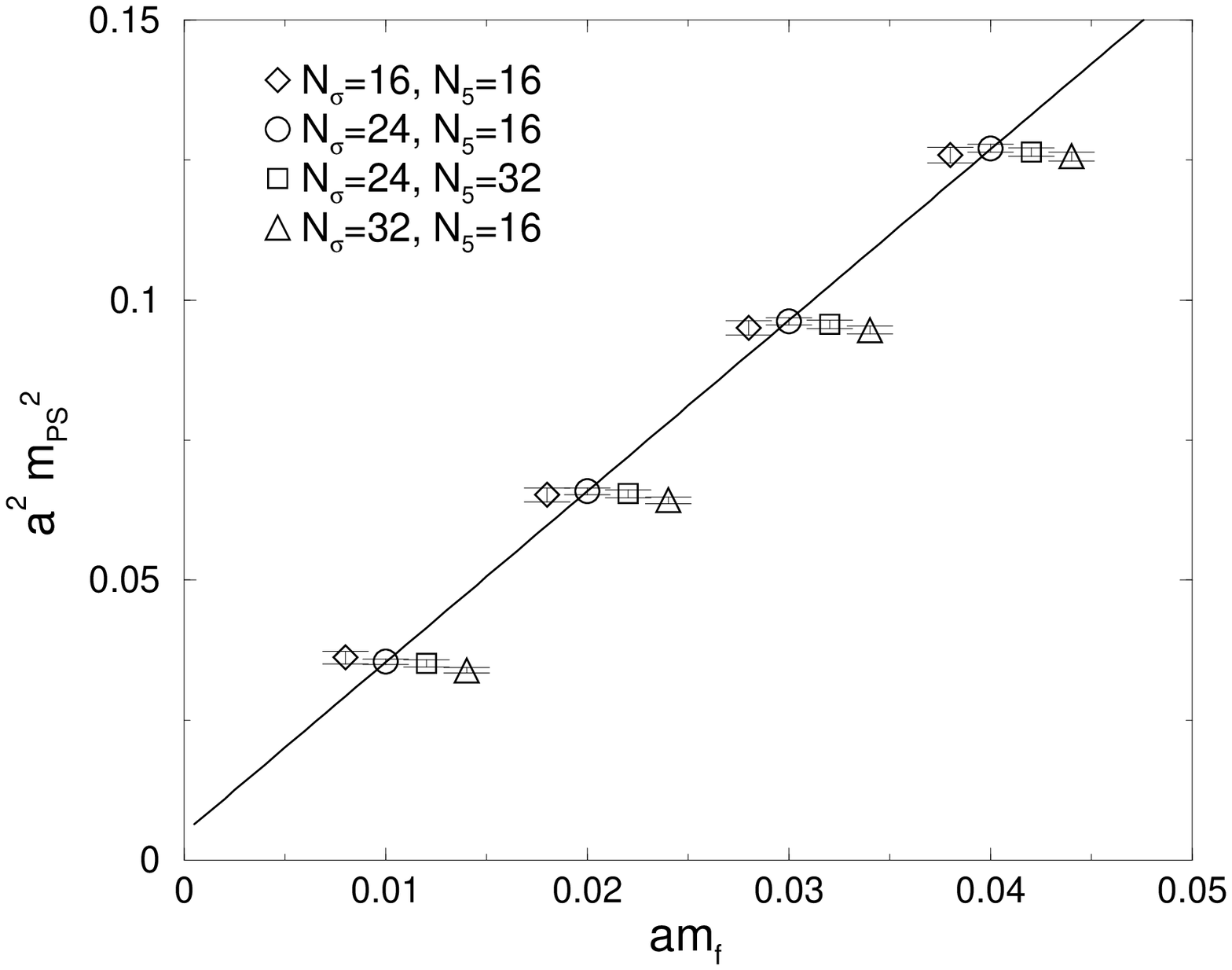}
  \epsfxsize=8cm \epsfbox{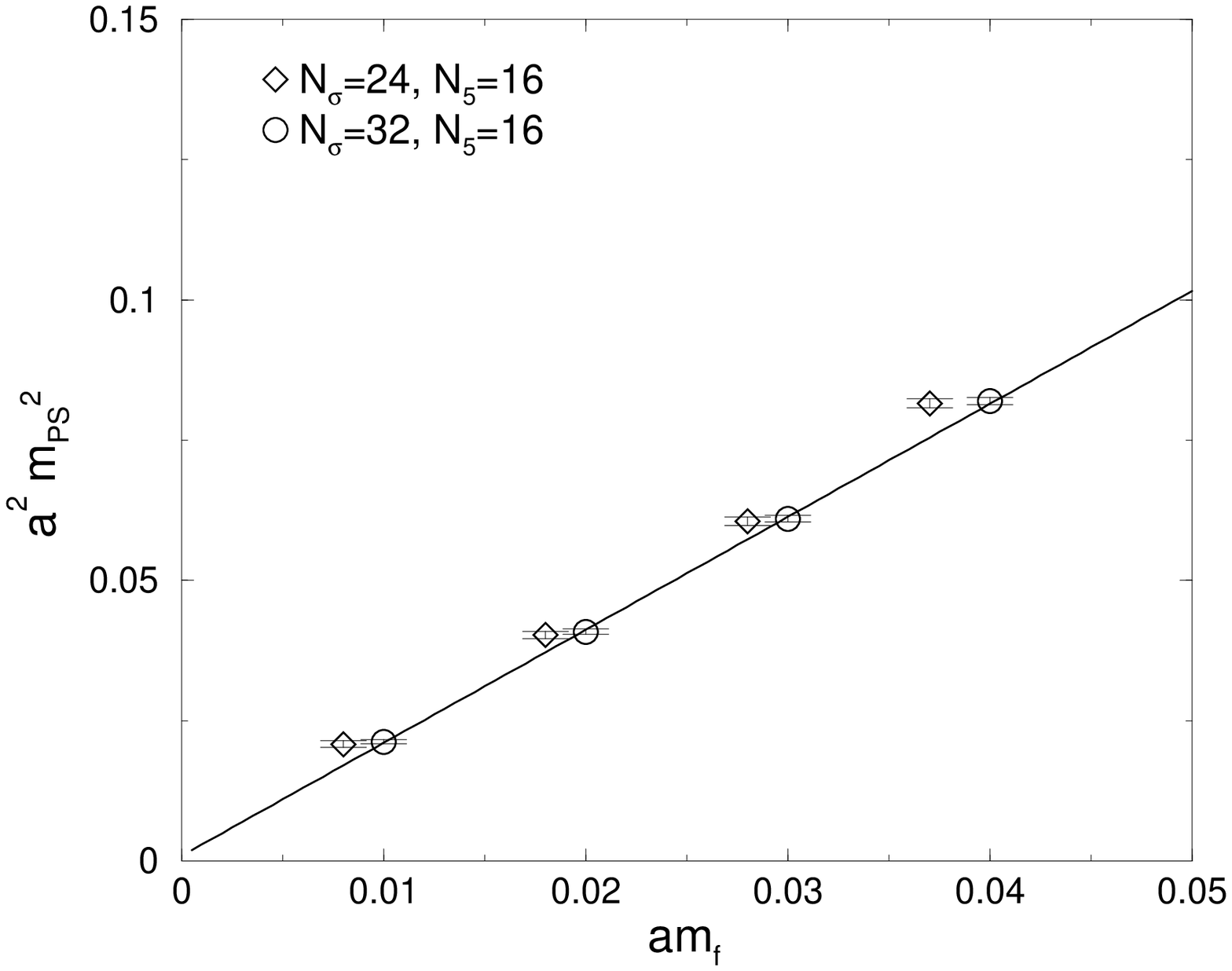}
  \caption{Pseudo scalar meson mass squared as a function of bare quark
  mass $m_f$ at $\beta=2.6$ (left) and at $\beta=2.9$ (right).
  Lines show linear fits to main runs.
  The data except for the main run are shifted in $m_f$.}
  \label{fig:mpi2-mf}
 \end{center}
\end{figure}

We have already mentioned that the pseudo scalar meson mass, if linearly
extrapolated, does not vanish at $m_f=0$.
We have also examined alternative fits including either 
a quadratic term, $(m_fa)^2$, or a quenched chiral logarithm term,
$m_fa \log (m_f a)$, in addition to the linear term.  
We have found that these yield almost identical values of the pseudo scalar 
meson mass at $m_f=0$.
We observe from the results at $\beta=2.6$ 
shown in the left panel of Fig.~\ref{fig:mpi2-mf}  
that the non-zero pseudo scalar meson mass cannot be explained as an
effect of finite fifth dimensional lengths, since the data at $N_5=16$
(open circles) and $N_5=32$ (open squares) are consistent within the
error down to the smallest quark mass $m_fa=0.01$.  
This conclusion is also supported by an analysis of the anomalous 
quark mass $m_{5q}$ defined by the axial Ward-Takahashi 
identity \cite{cppacs-dwf}.  This quantity provides a measure of chiral 
symmetry breaking due to a finite $N_5$.  It was found that $m_{5q}$ has 
only a very small value of 
$m_{5q}=0.274(42)$~MeV for $N_5=16$ at $\beta=2.6$. For comparison, 
the magnitude of $m_{res}$ obtained from the linear fit is $2-4$~MeV
as one can see from Table \ref{tab:massresults}.  

Examining the spatial size dependence of results at $\beta=2.6$ 
(left panel of Fig.~\ref{fig:mpi2-mf}) for 
$N_\sigma=16, 24$ and 32, we observe that the three points are mutually
consistent within the errors for the heavier quark mass of $m_fa=0.04, 0.03$ 
and 0.02, but that they show a decrease toward larger spatial volumes at
our lightest quark mass $m_fa=0.01$.  This indicates that    
the non-zero pseudo scalar meson mass at $m_f=0$ in the linear
extrapolation reflects a finite spatial volume effects in our pseudo scalar
meson mass data. 

\begin{figure}[p]
 \begin{center}
  \leavevmode
  \epsfxsize=9cm \epsfbox{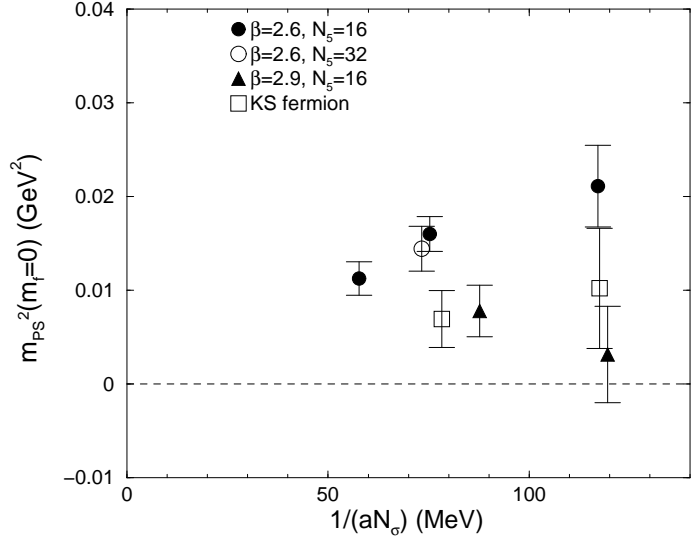}
  \caption{Pseudo scalar meson mass at $m_f=0$ as a function of
  spatial lattice size $N_\sigma$.
  Filled symbols represent data at fifth dimensional length $N_5=16$
  and an open circle represents that at $N_5=32$.
  The results of the KS fermion at similar volume size
  \protect\cite{AU94} is also plotted with open squares for comparison.}
  \label{fig:mpi2-V}
 \end{center}
\end{figure}

To make this point explicit, we plot the values of $m_{PS}^2$ at $m_f=0$ 
as a function $1/N_\sigma a$ in Fig.~\ref{fig:mpi2-V}.  
At $\beta=2.6$ the results (filled circles) exhibit a decrease as  
$1/N_\sigma a\to 0$. 
For comparison we plot by open squares results for the Kogut-Susskind 
quark action, which retains $U(1)$ chiral symmetry, 
obtained at a similar lattice spacing of $a^{-1}\approx 2$~GeV and 
spatial lattice sizes of $N_\sigma\approx 16-24$ \cite{AU94}. 
A similar magnitude of $m_{PS}^2$ in the chiral limit between the two quark 
actions both having chiral symmetry corroborates finite-size effects as the 
origin of non-zero values $m_{PS}^2$.   

The two points for $\beta=2.9$ do not show a clear volume dependence.  
This reflects an absence of spatial size dependence at $m_fa=0.01-0.04$ 
observed in the right panel of Fig.~\ref{fig:mpi2-mf}.  Quark 
masses in this range are heavier than those at $\beta=2.6$ due to a 
smaller lattice spacing, and hence calculations at smaller values of $m_fa$ 
are needed to expose finite spatial volume effects at $\beta=2.9$.

\section{$B$ parameters}
\label{sec:BK}

\subsection{Extraction of $B$ parameters}

\begin{figure}[p]
 \begin{center}
  \leavevmode
  \epsfxsize=8cm \epsfbox{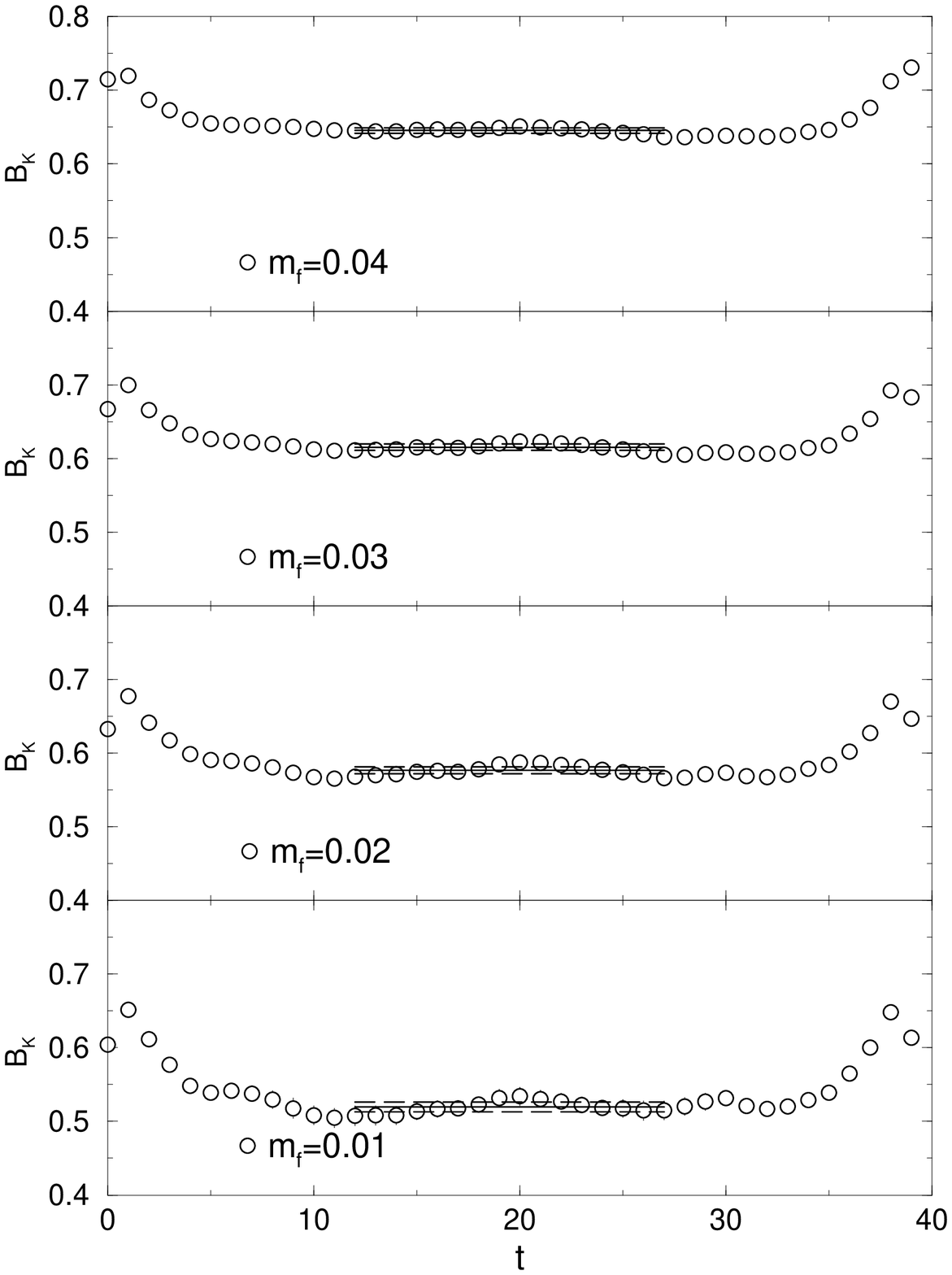}
  \epsfxsize=8cm \epsfbox{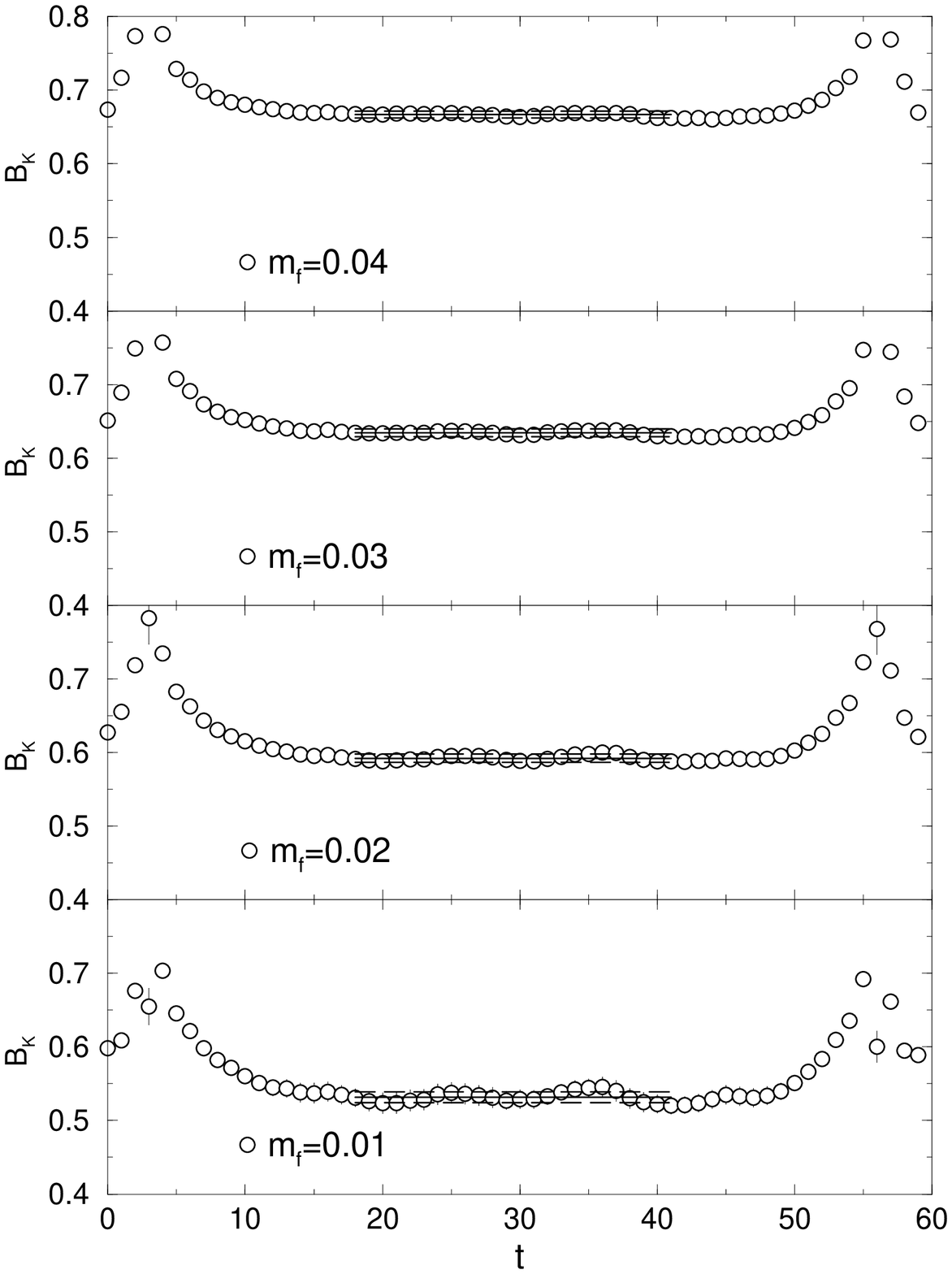}
  \caption{Ratio of weak matrix element with vacuum saturation
  \eqn{eqn:BK} as a function of temporal distance $t$ at $\beta=2.6$ on
  a $24^3\times40\times16$ lattice (left) and at $\beta=2.9$ on a
  $32^3\times60\times16$ lattice (right).
  Lines show constant fit over the fitted range.}
  \label{fig:BK-t}
 \end{center}
\end{figure}

\begin{figure}[p]
 \begin{center}
  \leavevmode
  \epsfxsize=8cm \epsfbox{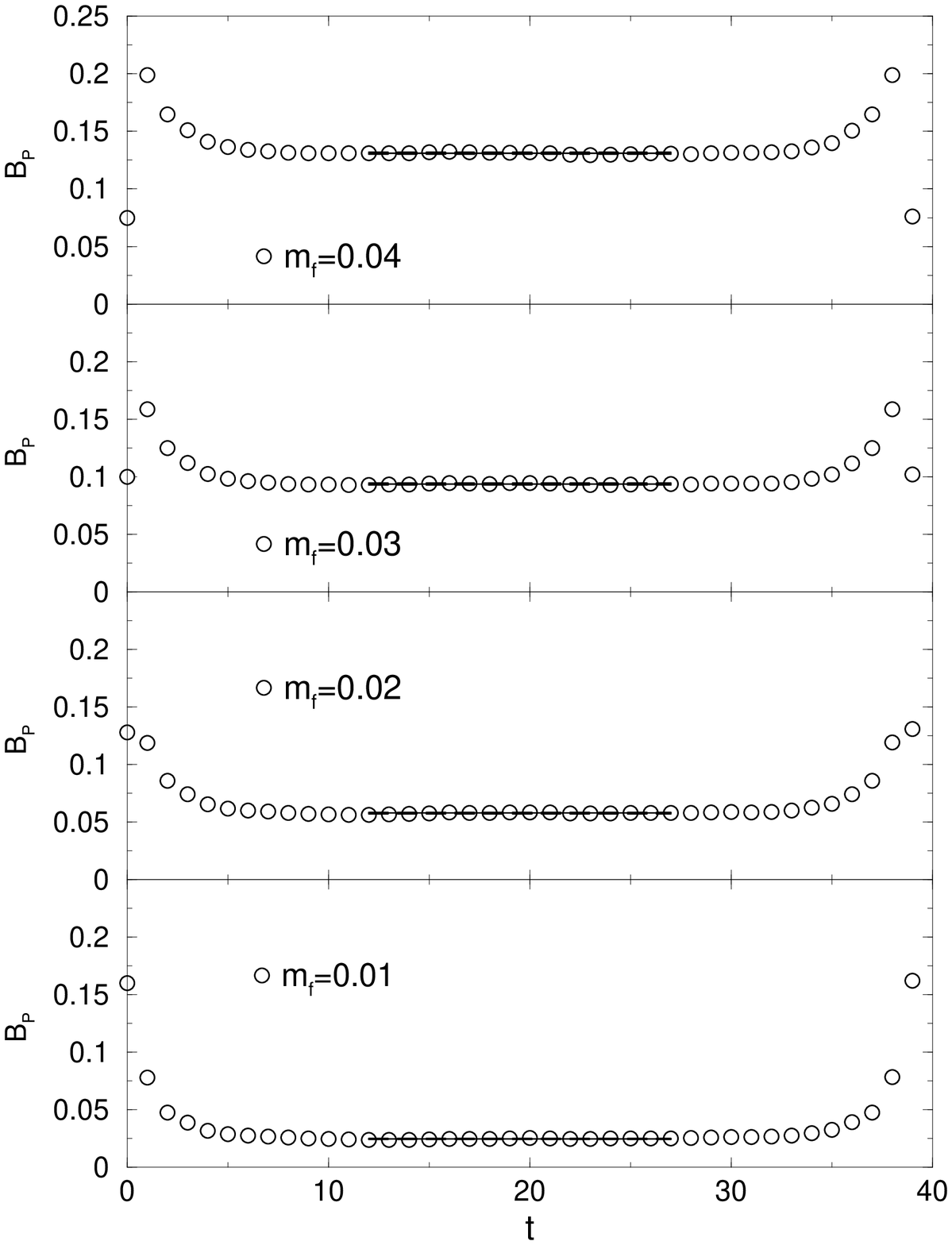}
  \epsfxsize=8cm \epsfbox{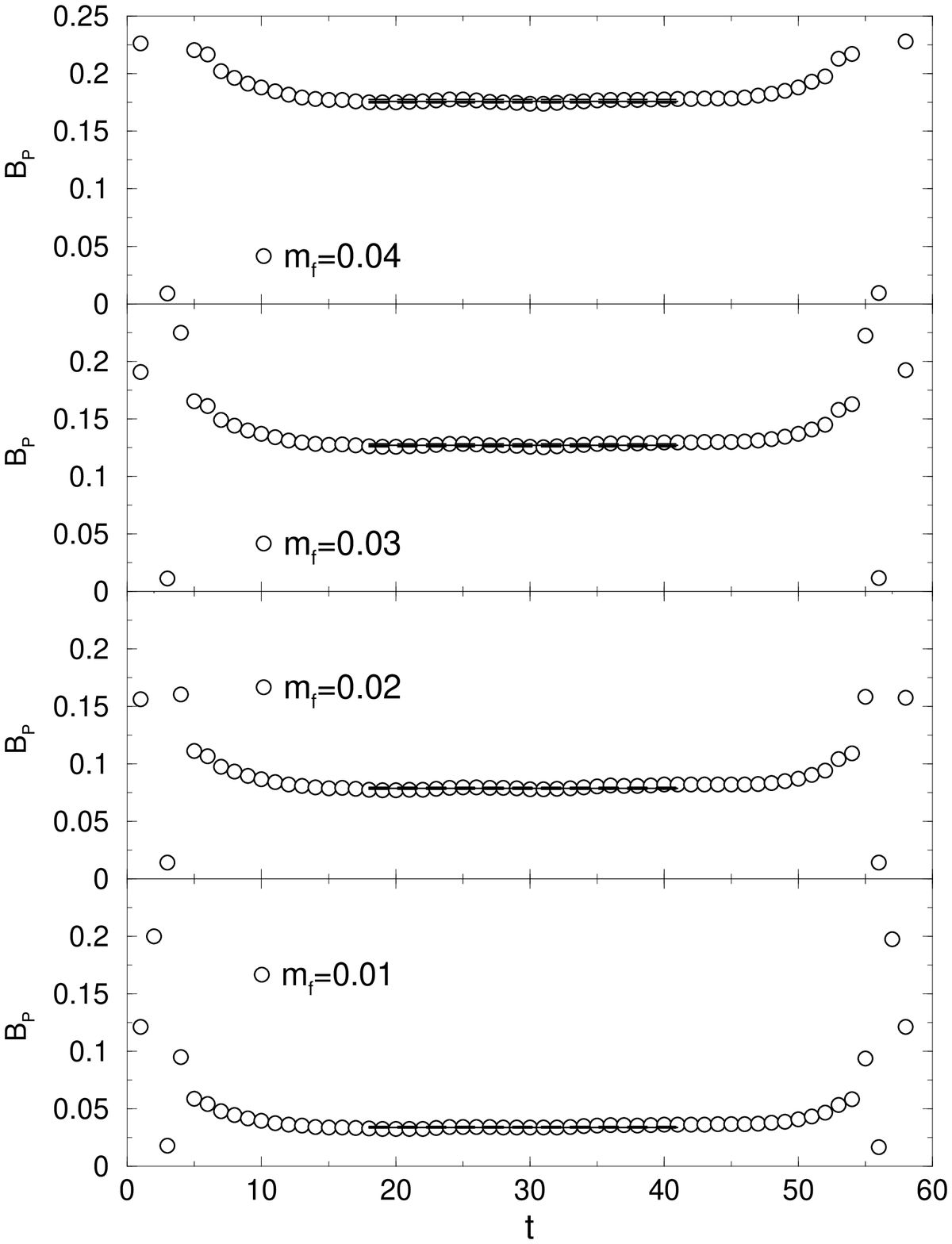}
  \caption{Ratio of weak matrix element with pseudo scalar density
  \eqn{eqn:BP} as a function of temporal distance $t$ at $\beta=2.6$ on
  a $24^3\times40\times16$ lattice (left) and at $\beta=2.9$ on a
  $32^3\times60\times16$ lattice (right).
  Lines show constant fit over the fitted range.}
  \label{fig:BP-t}
 \end{center}
\end{figure}

In Fig.~\ref{fig:BK-t} and \ref{fig:BP-t} we show typical data
for the ratio of kaon Green functions 
for $B_K$ and $B_P$ defined in \eqn{eqn:BK} 
and \eqn{eqn:BP} as a function of the temporal site $t$ of the weak 
operator. The values of these quantities at each $m_f$,
$N_\sigma$ and $N_5$ are extracted by fitting the plateau with a constant.  
The fitting range, determined by the inspection of plots
for the ratio and those for the effective pseudo scalar meson mass, is 
$12\le t \le 27$ for all simulations at $\beta=2.6$ and 
$18\le t \le 41$ at $\beta=2.9$.
In Tables \ref{tab:R26.16x40x16}-\ref{tab:R29.32x60x16} we list the
numerical values of $B_K$ and $B_P$ at four quark masses
$m_fa=0.01, 0.02, 0.03, 0.04$ for each set of run parameter.

\subsection{Chiral property for $B_P$}

\begin{figure}[p]
 \begin{center}
  \leavevmode
  \epsfxsize=8cm \epsfbox{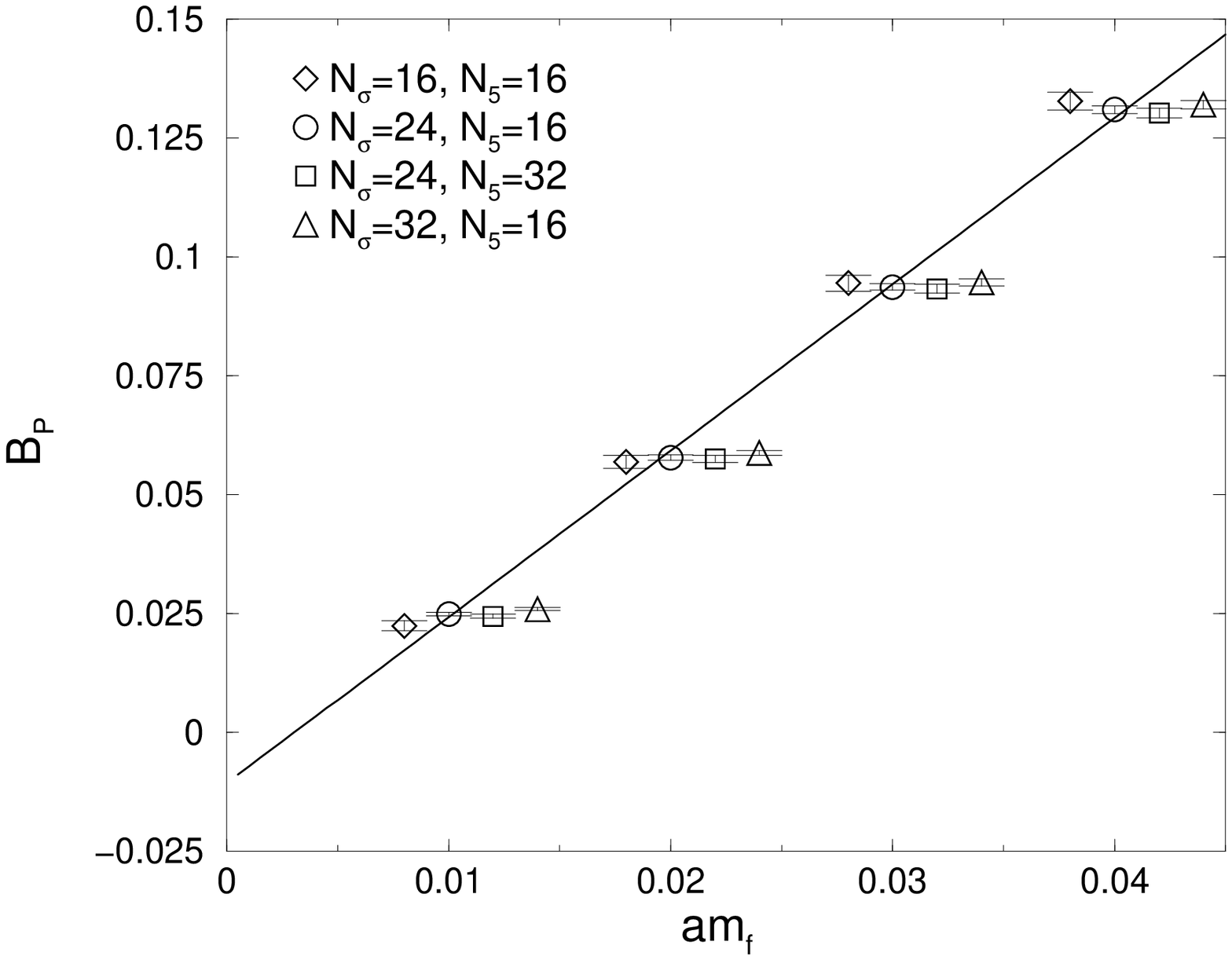}
  \epsfxsize=8cm \epsfbox{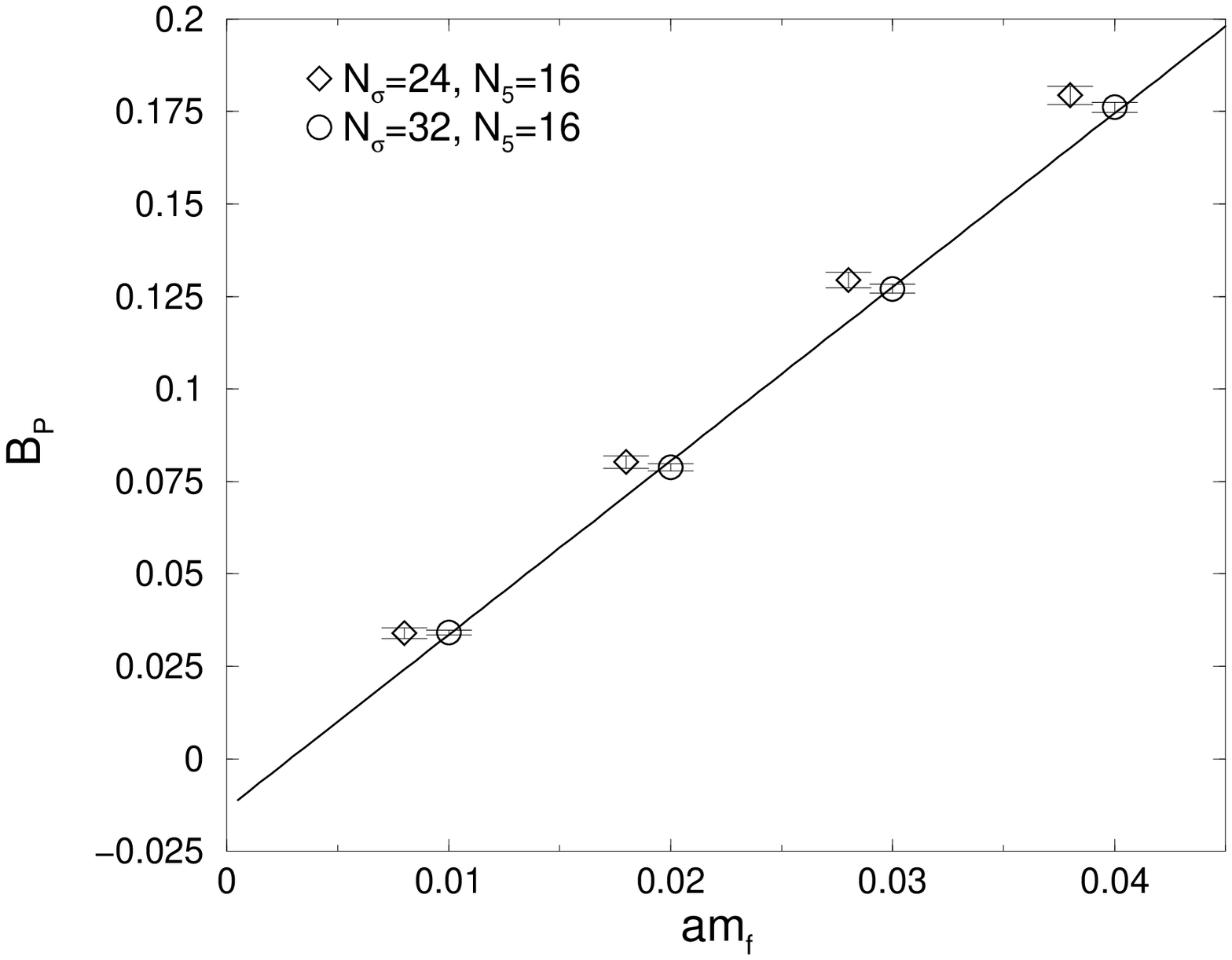}
  \caption{$B_P(m_f=0)$ vs $m_f$ at $\beta=2.6$ (left) and $\beta=2.9$
  (right).
  The data except for the main run are shifted in $m_f$.}
\label{fig:BP-mf}
 \end{center}
\end{figure}

We have seen in Sec.~\ref{sec:hadron} that corrections in $m_{PS}^2$ due
to a finite fifth dimensional size $N_5$ is sufficiently small for
$N_5=16$ for the range of quark mass $m_f$ explored, and that
the non-zero pseudo scalar meson mass at $m_f=0$ is caused by finite
spatial size effects.
As a further check
we investigate the chiral property of the matrix element for the
four-quark operator through $B_P$, which is expected to vanish linearly 
at $m_f=0$.
In Fig.~\ref{fig:BP-mf} we plot bare values of $B_P$ as a function of
$m_fa$ at $\beta=2.6$ and $2.9$.
Inspecting the results at $\beta=2.6$ on the left panel of 
Fig.~\ref{fig:BP-mf} we observe an agreement for the fifth dimensional size 
$N_5=16$ (open circle) and 32 (open square).  
This shows that $N_5=16$ is also large enough for this matrix element.

On the other hand, there is a
trend of increase for larger spatial volumes when the quark mass goes 
below $m_f=0.02$.  Finite spatial size effects appear also
in this quantity.
Making a linear chiral extrapolation, we find a small 
but negative residual at $m_f=0$.
Contrary to the case of $m_{PS}^2$,
two alternative fits including a quadratic term, $(m_f a)^2$, or
a chiral logarithm term, $m_f a\ \log(m_f a)$,
give smaller sizes of the intercept at $m_f=0$ compared to
that from the linear fit.
We find, however, that sizes of the intercept
decrease as $N_\sigma$ increases for all fits.
Therefore we conclude that the non-zero values of $B_P$ at $m_f=0$ are
a finite-spatial size effect, and is not a signal of violation
of chiral symmetry.

The negative sign of the intercept may be
understood as follows.
Chiral symmetry implies
\begin{eqnarray}
Z_A
\frac{\langle 0 \vert A_\mu \vert P\rangle}{\langle 0 \vert P \vert P\rangle}
&=& \frac{2 m_f}{m_{PS}},
\end{eqnarray}
where the bare quantities $A_\mu$ and $P$ are local axial vector current 
and pseudo scalar density, and $Z_A$ is 
the renormalization factor for $A_\mu$, with which we obtain
\begin{eqnarray}
B_P &=& \frac{8}{3}B_K 
\frac{\vert\langle 0 \vert A_\mu \vert P\rangle\vert^2}
{\vert\langle 0 \vert P \vert P\rangle\vert^2}
= B_K \frac{32 m_f^2}{3 Z_A^2 m_{PS}^2 } 
= B_K \frac{32 m_f^2}{3 Z_A^2 A_{PS}(m_f + m_{res}) }.
\end{eqnarray}
This relation is well-satisfied
at $\beta = 2.6$ where $Z_A$ is non-perturbatively known\cite{Saoki}
and
is reasonably good with the perturbative $Z_A$ at $\beta=2.9$.
Since $ m_{res} \ll m_f$ in the range of $m_f$ in our simulation,
we approximately obtain
\begin{eqnarray}
B_P &\simeq & B_K\frac{32}{3 Z_A A_{PS}} ( m_f -m_{res} ),
\end{eqnarray}
showing that a positive $m_{res}$ implies a negative intercept of $B_{P}$.
This formula also suggests
that the large part of the size effect for $B_P$ is
caused by that for $m_{PS}^2$.

\subsection{$B_K$}
\label{sec:bk}

\begin{figure}[tb]
 \begin{center}
  \leavevmode
  \epsfxsize=8cm \epsfbox{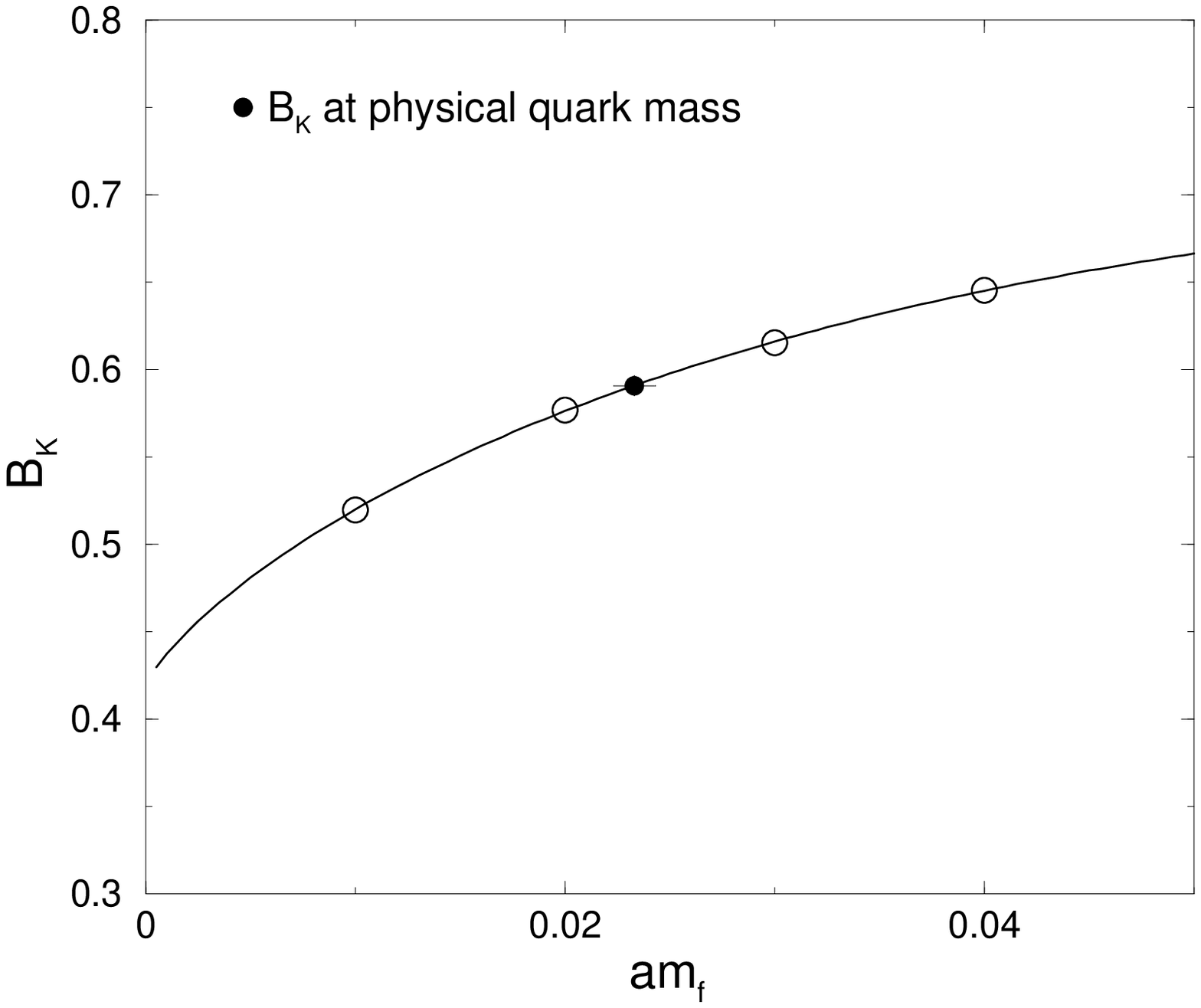}
  \epsfxsize=8cm \epsfbox{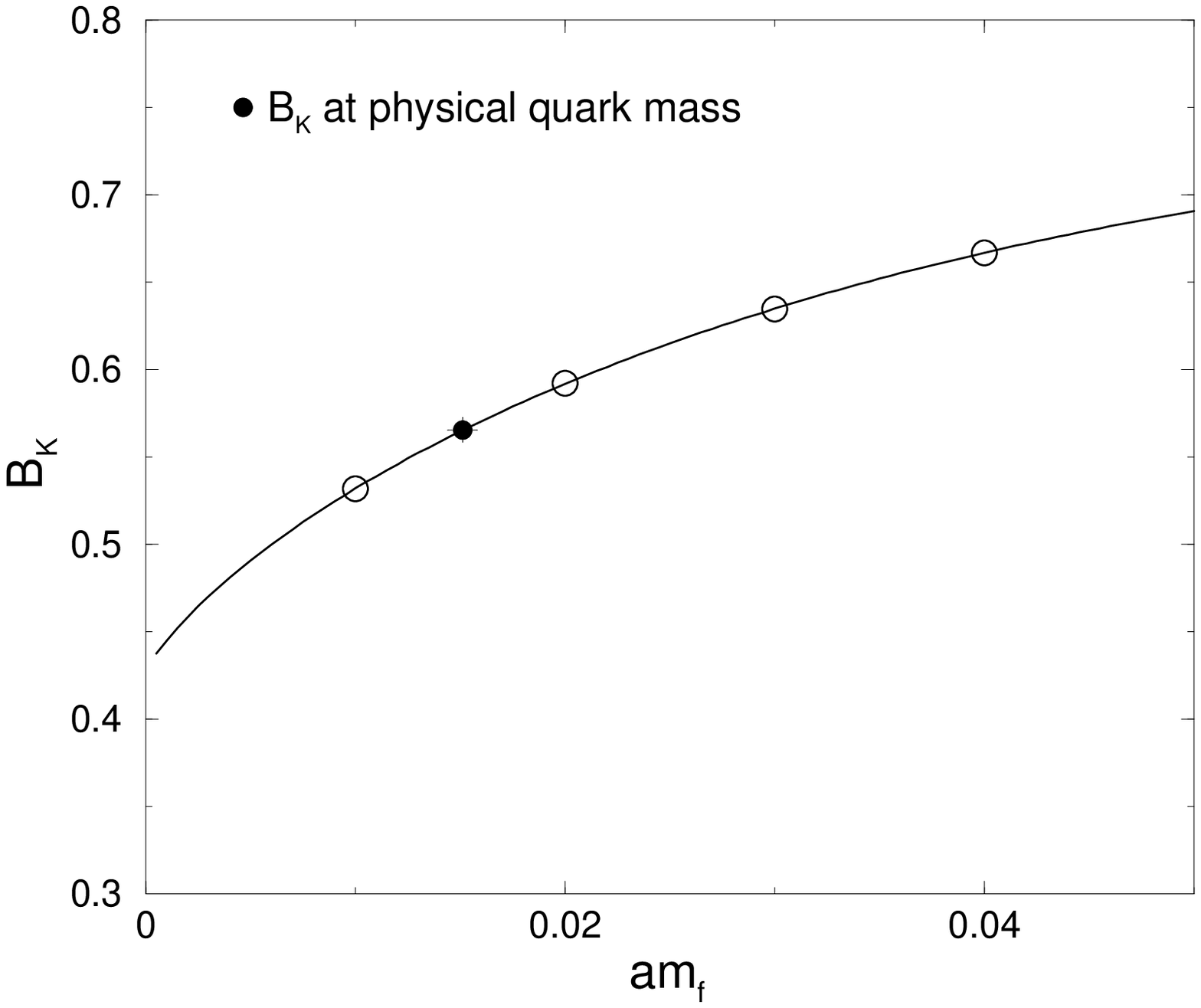}
  \caption{Bare $B_K$ interpolated as a function of 
  $m_fa$ at $\beta=2.6$ for a $24^3\times40\times16$ lattice (left) and
  at $\beta=2.9$ on a $32^3\times60\times16$ lattice (right).}
\label{fig:BK-mf}
\end{center}
\end{figure}

The bare value of $B_K$ is interpolated as a function of $m_fa$ using a 
formula suggested by chiral perturbation theory\cite{Sharpe92},
\begin{eqnarray}
B_K=B\left(1-3c m_fa\log(m_fa) + b m_fa\right).
\end{eqnarray}
This interpolation is illustrated in Fig.~\ref{fig:BK-mf}. 
The physical value of $B_K$ is obtained at the point $m_f=m_f^s/2$ 
(solid circles in Fig.~\ref{fig:BK-mf}) 
which is estimated from the experimental value of $m_K/m_\rho$.
The renormalized values of $B_K({\rm NDR};\mu=2 {\rm GeV})$ and related 
physical quantities are collected 
in Table~\ref{tab:results}.

\begin{figure}[tb]
 \begin{center}
  \leavevmode
  \epsfxsize=9cm \epsfbox{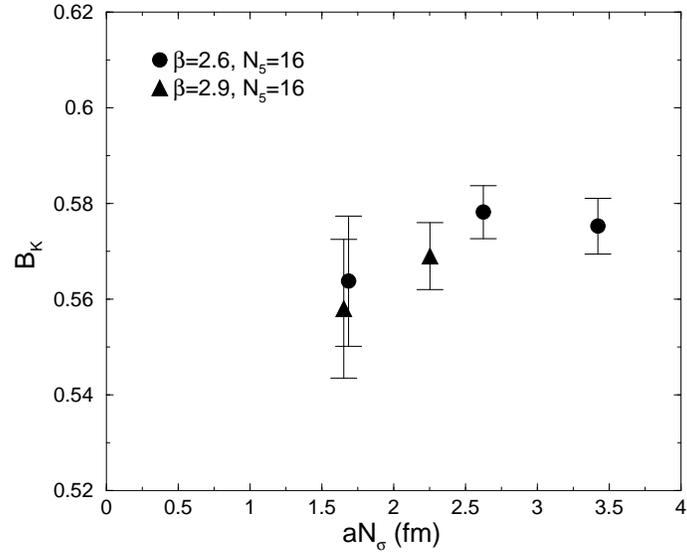}
\caption{Renormalized $B_K$ as a function of spatial size.}
\label{fig:BK-V}
 \end{center}
\end{figure}

We plot the renormalized value of $B_K$
as a function of the spatial size in Fig.~\ref{fig:BK-V}.
Filled circles and triangles are results at $\beta=2.6$ and $2.9$ 
keeping the same fifth dimensional size $N_5=16$.
At $\beta=2.6$ we observe a slight increase of $B_K$ from the spatial size 
$N_\sigma a\approx 1.7$~fm to 2.6~fm, but the values beyond the size 
$N_\sigma a\approx 2.6$~fm are well consistent within the statistical 
error of 1\%.  
This result agrees with that of a previous finite spatial size study 
with the Kogut-Susskind quark action \cite{jlqcd}, which found
finite size effects to be smaller than 0.5\% for the spatial size 
$N_\sigma a\gsim 2.2$~fm. 
We conclude that the size of about 2.6~fm ($N_\sigma =24, \beta=2.6$) 
and 2.3~fm ($N_\sigma=32, \beta=2.9$) used in our main runs is
sufficient to avoid spatial size effects for $B_K$ at a 1\% level.

\begin{figure}[tb]
 \begin{center}
  \leavevmode
  \epsfxsize=9cm \epsfbox{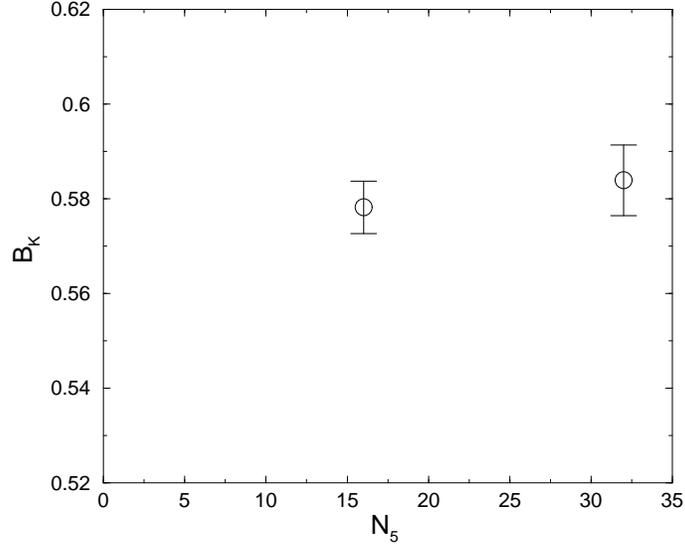}
\caption{Renormalized $B_K$ as a function of fifth dimensional length
  $N_5$.}
\label{fig:BK-N}
 \end{center}
\end{figure}

In Fig.~\ref{fig:BK-N} we plot $B_K$ as a function of the fifth
dimensional length $N_5$
on a $24^3\times40$ four-dimensional lattice at $\beta=2.6$. 
The results at $N_5=32$ and $N_5=16$ are in agreement within
the statistical error of 1\%.  Hence the fifth dimensional size of
$N_5=16$ is sufficient for the calculation of $B_K$ at this accuracy. 

\begin{figure}[tb]
 \begin{center}
  \leavevmode
  \epsfxsize=9cm \epsfbox{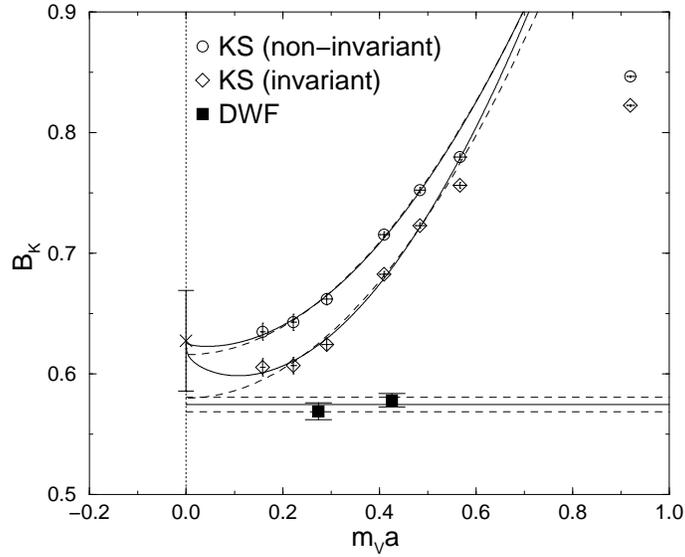}
\caption{Scaling behavior of renormalized $B_K(\mu=2 {\rm GeV})$.
Previous results with the KS action\protect\cite{jlqcd} are also shown
  with open symbols.}
\label{fig:BK-a}
 \end{center}
\end{figure}

Our final results from the main runs are shown in Fig.~\ref{fig:BK-a}
as a function of lattice spacing by filled squares.  
The open symbols and the associated lines represent results from a 
previous calculation with the Kogut-Susskind (staggered) 
quark action \cite{jlqcd}, where
gauge invariant and non-invariant four-quark operators are used.
Our result obtained with the domain wall quark action 
and an RG-improved gluon action show a much better scaling behavior;
the central values of the two points differ by only 1.6\% while 
the Kogut-Susskind results show a 10\% decrease over the similar range 
of lattice spacing $a^{-1}\approx 2-3$~GeV. 
In order to estimate the continuum value, we then make a constant 
extrapolation $B_K(a)=B_K$, which yields $B_K(\mu=2\mbox{GeV})= 0.5746(61)$. 

Possible sources of systematic errors in this result are 
scaling violation ignored in the constant fit and 
higher loop corrections in the renormalization factors.
Making an extrapolation of our data of the form $B_K(a)=B_K+c\cdot a^2$,
based on ${\cal O}(a^2)$ scaling violation expected for DWQCD
\cite{BSW99,TN99},
we obtain an estimate of 2.2\% for the first error.    
A simple estimate for the second error is provided by 
the value of $\alpha_{\overline{\rm MS}}(1/a)^2$ at the finer lattice spacing 
of $\beta=2.9$.  This yields 2.5\% for the second error.  
This seems to be a reasonable estimate 
since other methods of estimation, either shifting the matching scale 
from $q^*=1/a$ to $q^*=\pi/a$ or employing different choices of gauge coupling 
such as \eqn{eqn:coupling-P}, give a small variation of ${\cal O}(1\%)$.
Adding the two estimates by quadrature 
 gives a 3.3\% systematic error, and we obtain 
\begin{equation}
B_K(\mbox{NDR}; \mu=2\mbox{GeV})= 0.5746(61)(191).
\end{equation}
as our estimate of the continuum value of $B_K$ in the $\overline{\rm MS}$ 
scheme at $\mu=2$~GeV.

This value lies at the lower edge of the one-standard deviation 
error band of the result $B_K(\mu=2\mbox{GeV})=0.628(42)$
obtained with the Kogut-Susskind action\cite{jlqcd}.
We recall that the statistical error with the Kogut-Susskind results are 
at the $0.5-1$\% level. A significantly larger error of 6.7\% in the
continuum value arises from the continuum extrapolation incorporating
both the $a^2$ scaling violation and the $\alpha_s^2$ uncertainty due to
the use of one-loop renormalization factor.  
Making a more detailed check of agreement of results from the two types of 
quark actions requires a better control of systematic errors,  
in particular those due to renormalization factors. 
For this purpose non-perturbative determination of these factors for 
both cases will be necessary. 

The RBC Collaboration carried out a quenched simulation with the domain-wall 
quark action and a plaquette gluon action at $\beta=6.0$ ($a^{-1}\approx 
2$~GeV) on a $16^3\times 32\times 16$ lattice.  Employing the 
method of Ref.~\cite{NPR} to non-perturbatively determine the 
renormalization factors, they reported a value 
$B_K(\mu=2\mbox{GeV})=0.538(8)$\cite{RBC}.   
This value is 7\% smaller than our 
result.  A precise comparison, however, would require examination of 
spatial size and scaling violation effects in the RBC result, and 
of renormalization factors in our result as discussed above. 

\subsection{$B_K$ as a function of $m_{PS}^2$ in the continuum limit}

We have so far discussed the scaling behavior
of $B_K$ at the physical quark mass. 
Our data, in fact, allows us to examine
the scaling behavior of $B_K$ over a wide range of quark mass,
and derive the mass dependence of $B_K$ in the continuum limit.

\begin{figure}[tb]
 \begin{center}
  \leavevmode
  \epsfxsize=9cm \epsfbox{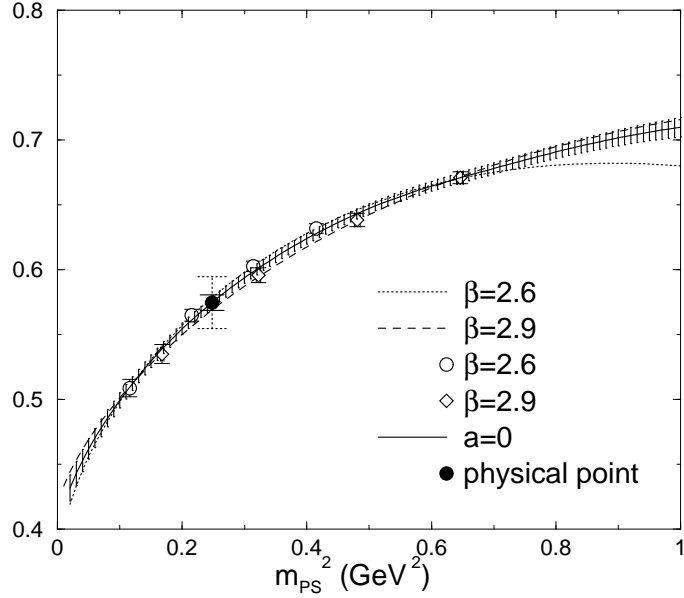}
\caption{Renormalized $B_K(\mu = 2 {\rm GeV})$ as a function of
$m_{PS}^2$, at $\beta = 2.6$, 2.9 and $\infty$(continuum limit),
where errors are shown only in the continuum limit.
Open symbols represent data obtained in our simulations, while the
solid circle gives the value of $B_K$ at the physical point in the
  continuum limit, with the statistical error(solid) and the total
  error(dotted).}
\label{fig:contbk}
 \end{center}
\end{figure}

In order to compare results at different lattice spacings, we employ
$m_{PS}^2$ in physical units (GeV$^2$) instead of $m_f a$.
In Fig.~\ref{fig:contbk} $B_K (\mu = 2 {\rm GeV})$ is given as a function of
$m_{PS}^2$ (GeV$^2$) at $\beta=2.6$,  $2.9$ and in the continuum limit.
The data are first fitted by
\begin{eqnarray}
B_K &=& B ( 1 - 3\ c\ m_{PS}^2\ \log(m_{PS}^2) + b\ m_{PS}^2 )
\label{eq:bkmpi}
\end{eqnarray}
for each $\beta$ and then extrapolated to the continuum by a constant
fit.
All errors in the figure are estimated by a single elimination
jackknife procedure, except for fit errors for the continuum extrapolation.
As seen in the figure, scaling violation is mild up to
$m_{PS}^2 \le 0.8$ GeV$^2$, and the continuum extrapolation is
reliable there. This confirms that the small scaling violation of
the physical $B_K$ observed in Sec.~\ref{sec:bk} is not an accidental one
at $m_{PS}=m_K$ but it holds over a wide range of the pseudo scalar
meson mass.

In Table~\ref{tab:contbk}, values of $B_K$ in the continuum limit,
which are also fitted by the same form (\ref{eq:bkmpi}),
are given for $ 0.02\le m_{PS}^2 \le 1.0$ (GeV$^2$) with errors.
Fitted parameters $B$, $b$ and $c$
are also given in the table, together with the reproduced values.
From this result in the continuum limit
one can see that
the contribution from higher order terms in chiral perturbation 
theory ($b$ and $c$) is non-negligible
and becomes as large as 40\% of the leading order contribution ($B$) at
$m_{PS} = m_K$.

\section{Light quark masses}
\label{sec:ms}

We attempt a determination of light quark masses $m_{ud}\equiv (m_u+m_d)/2$ 
and $m_s$ using our meson mass data.  There is a difficulty 
associated with a non-zero pseudo scalar meson mass at $m_f=0$ due to 
finite spatial sizes, which is represented by $m_{res}$ in the linear 
chiral formula (\ref{eq:pion}).  This causes systematic uncertainties 
in the results for quark masses, which is quite sizable for light $u$ and $d$ 
quarks.

In order to examine this problem, 
we calculate the physical quark masses in two ways which differ in the 
choice of origin for bare quark mass.  
In the first method we take $m_f=0$ as the origin, and write 
\begin{eqnarray}
m_{ud}&=&Z_q m_f^{ud},\\
m_s   &=&Z_q (m_f^s-m_f^{ud}).
\end{eqnarray}
Here $m_f^{ud}$ and $m_f^s$ are the bare quark mass $m_f$  
for the physical point of pion and kaon
determined by (\ref{eq:udquark}) and (\ref{eq:squark}).  
The subtraction of $m_f^{ud}$ in the second equation is to take into 
account the contribution of $u$-$d$ quark in the kaon mass,
$m_K\propto m_{ud}+m_s$, and $Z_q$ denotes the renormalization factor 
to match the bare lattice value to that in the continuum in the 
$\msbar$ scheme with NDR at $\mu=2$~GeV as discussed in 
Sec.~\ref{sec:renormalization}. 

In the second case we take the point $m_f=-m_{res}$, 
where pseudo scalar meson mass vanishes, as the origin. The formula then reads
\begin{eqnarray}
m_{ud}&=&Z_q (m_f^{ud}+m_{res}),\\
m_s   &=&Z_q (m_f^s-m_f^{ud}+m_{res}).
\end{eqnarray}
On the other hand the strange quark mass with the phi meson mass as input 
is given directly as
\begin{eqnarray}
m_s = Z_q m_s(\phi).
\end{eqnarray}
The results of these calculations are listed in Table \ref{tab:qmresults}.

\begin{figure}[tb]
 \begin{center}
  \leavevmode
  \epsfxsize=8cm \epsfbox{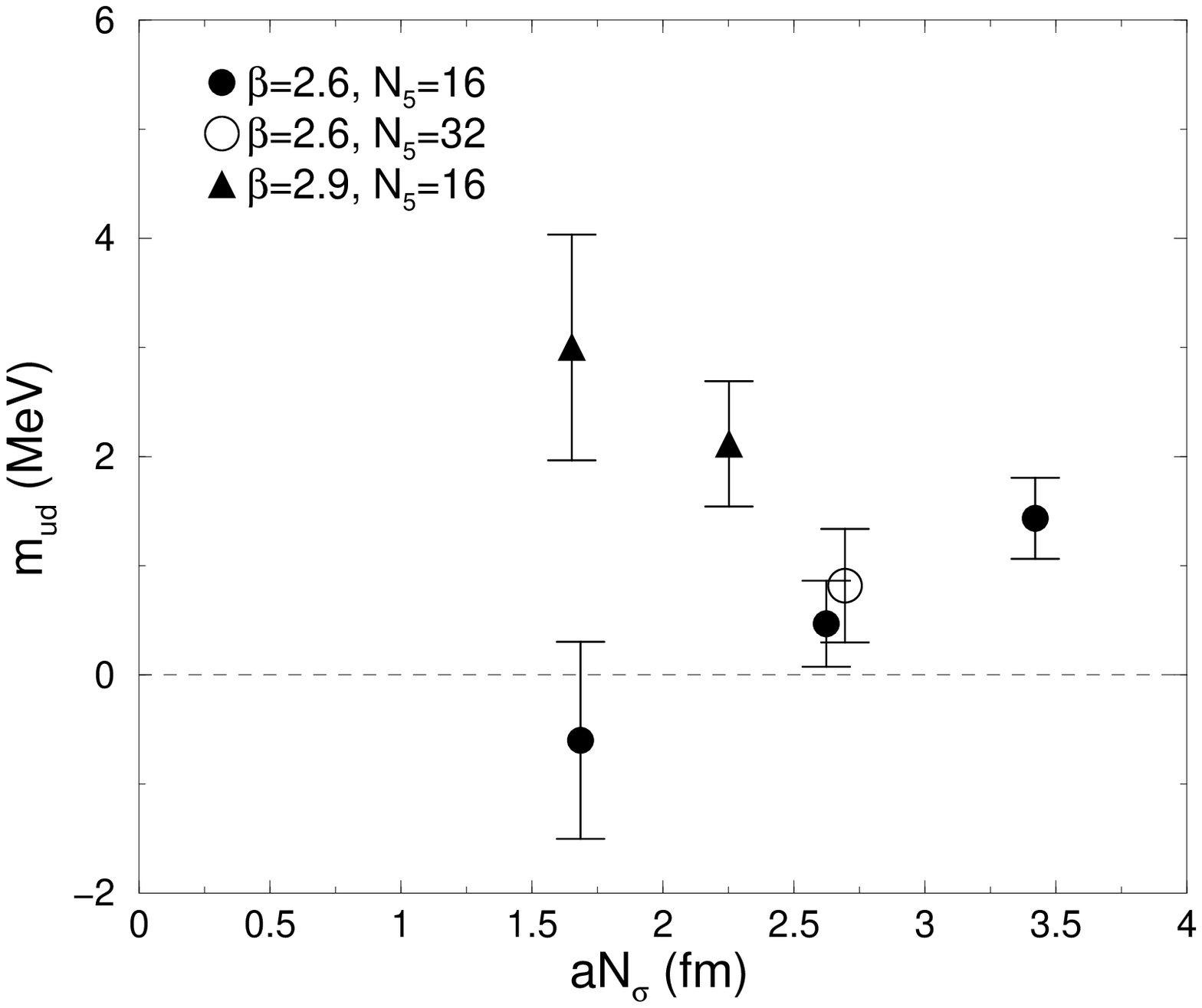}
  \epsfxsize=8cm \epsfbox{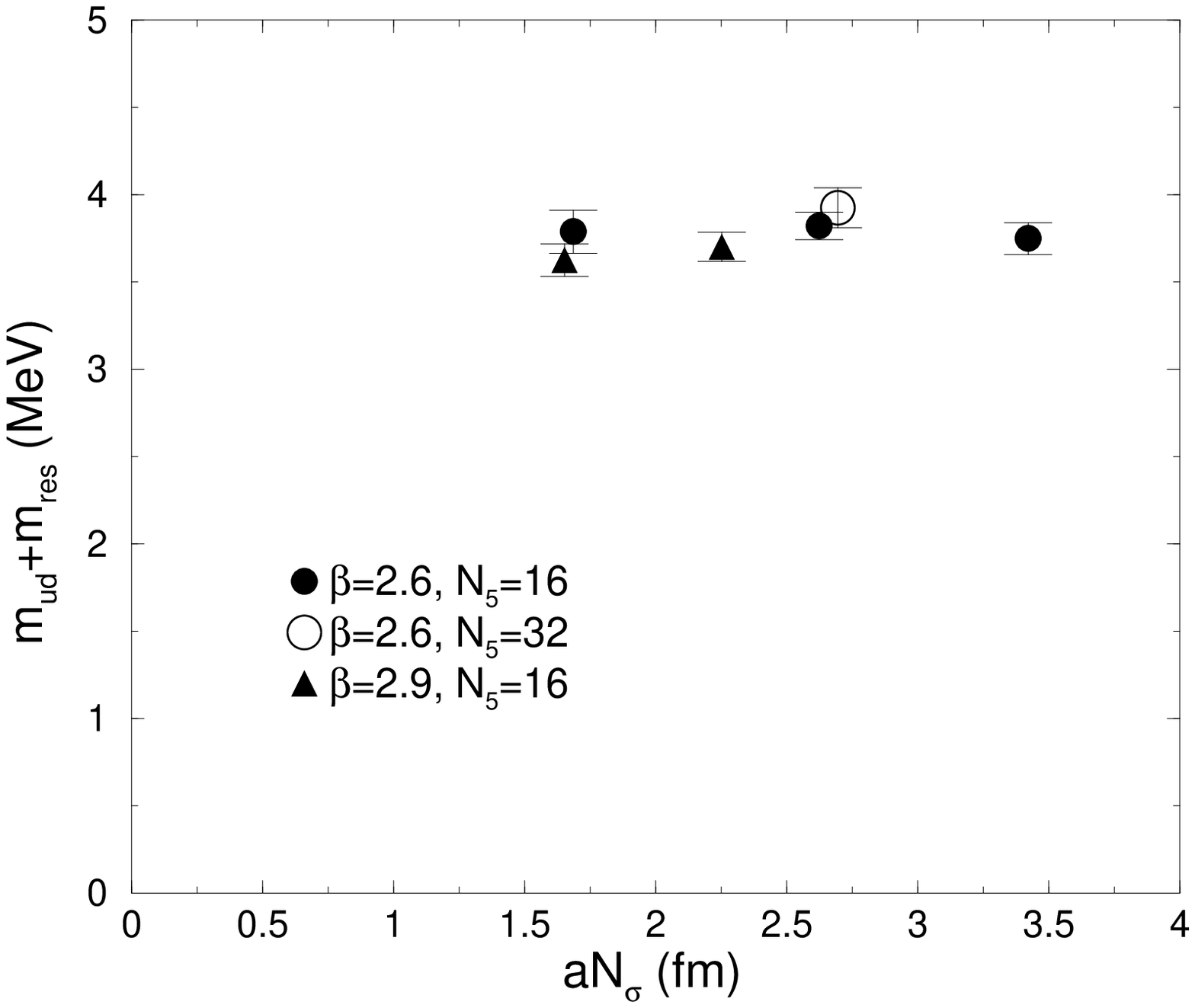}
\caption{Renormalized $u$-$d$ quark mass as a function of spatial size. 
  Results with (right) and without $m_{\rm res}$ added (left) are shown.}
\label{fig:ml-V}
 \end{center}
\end{figure}

In Fig.~\ref{fig:ml-V} we plot the $u$-$d$ quark mass calculated in the
two ways above as a function of spatial size $aN_\sigma$ in physical units. 
Two features are quite evident from this figure: 
(i) There is little dependence on the fifth dimensional size $N_5$. 
Hence $N_5=16$ is sufficient to avoid effects of chiral symmetry breaking 
at our range of lattice spacings. 
(ii) Effects of finite spatial size, by contrast, are quite significant if 
$m_{res}$ is ignored, even yielding a negative value for $m_{ud}$ 
for small spatial sizes (left panel of Fig.~\ref{fig:ml-V}).  
The values calculated including $m_{res}$, 
on the other hand, are much more stable as a function of $aN_\sigma$ 
(right panel).  

In order to understand the second point, 
we note that $m_{res}a$ depends strongly on the volume
while the slope $A_{PS}$ is almost volume independent.
Using (\ref{eq:udquark}) and the corresponding one at $N_\sigma
=\infty$ given by
\begin{equation}
\frac{\sqrt{ A_{PS}(N_\sigma =\infty)\cdot
m_f^{ud}(N_\sigma =\infty)a}}
{A_V+B_Vm_f^{ud}a}
=\frac{m_\pi}{m_\rho},
\end{equation}
and 
neglecting a small volume dependence of the denominator $A_V+B_V m_f^{ud}a$,
we observe that the following formula holds:
\begin{equation}
\frac{m_f^{ud}-m_f^{ud}(N_\sigma=\infty)}{m_f^{ud}(N_\sigma=\infty)}
=-\frac{m_{res}}{m_f^{ud}(N_\sigma=\infty)}
-\frac{A_{PS}-A_{PS}(N_\sigma=\infty)}{A_{PS}}.
\end{equation}
Since the magnitude $m_{res}\approx 2-3$~MeV for our spatial size of 
$aN_\sigma\approx 2.5$~fm is comparable to the actual $u$-$d$ quark mass, 
the first term is ${\cal O}(1)$ and becomes the main contribution
to the size effect, while the second term, 
representing finite size effect in the slope $A_{PS}$, is found to be
much smaller.
Hence including $m_{res}$ removes a dominant part of 
finite size effects in $u$-$d$ quark mass.  In view of this situation we 
take the values including $m_{res}$ as the best estimate from our present 
data for $m_{ud}$.

\begin{figure}[tb]
 \begin{center}
  \leavevmode
  \epsfxsize=9cm \epsfbox{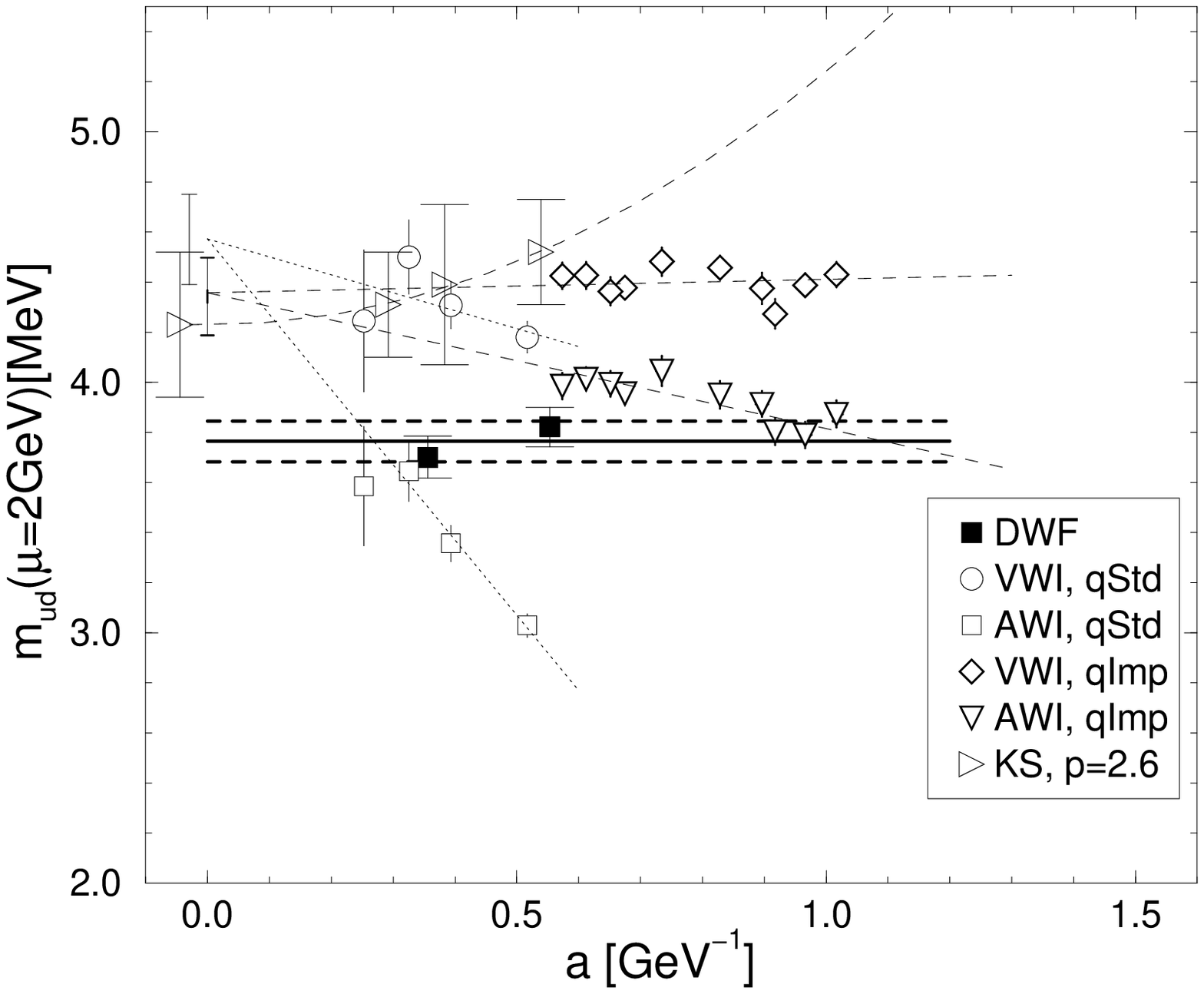}
\caption{
Scaling behavior of $u$-$d$ quark mass, calculated with $m_{\rm res}$
  added, compared with those from 4-dimensional quark action: 
Wilson action (Std) \protect\cite{cppacs-quenched},  
clover-improved action (Imp)\protect\cite{cppacs-full}, 
and Kogut-Susskind (KS) \protect\cite{jlqcd-ksqm} actions.
VWI and AWI represent vector and axial-vector Ward-Takahashi identity masses,
respectively.
$p$ for the KS fermion represents the matching scale of the RI scheme in 
  units of GeV.}
\label{fig:ml-a}
 \end{center}
\end{figure}

The scaling behavior of $m_{ud}$ is plotted in Fig.~\ref{fig:ml-a} 
by filled squares.  Making a constant fit to the two values, we find 
\begin{equation}
m_{ud}^{\overline{\rm MS}}(2 {\rm MeV})=3.764(81)(215) {\rm MeV},
\end{equation} 
where the first error is statistical and the second due to 
scaling violation and ${\cal O}(\alpha^2)$ systematic errors 
estimated in the same way as for $B_K$ (see Sec.~\ref{sec:bk})
and added in quadrature.  

\begin{figure}[tb]
 \begin{center}
  \leavevmode
  \epsfxsize=9cm \epsfbox{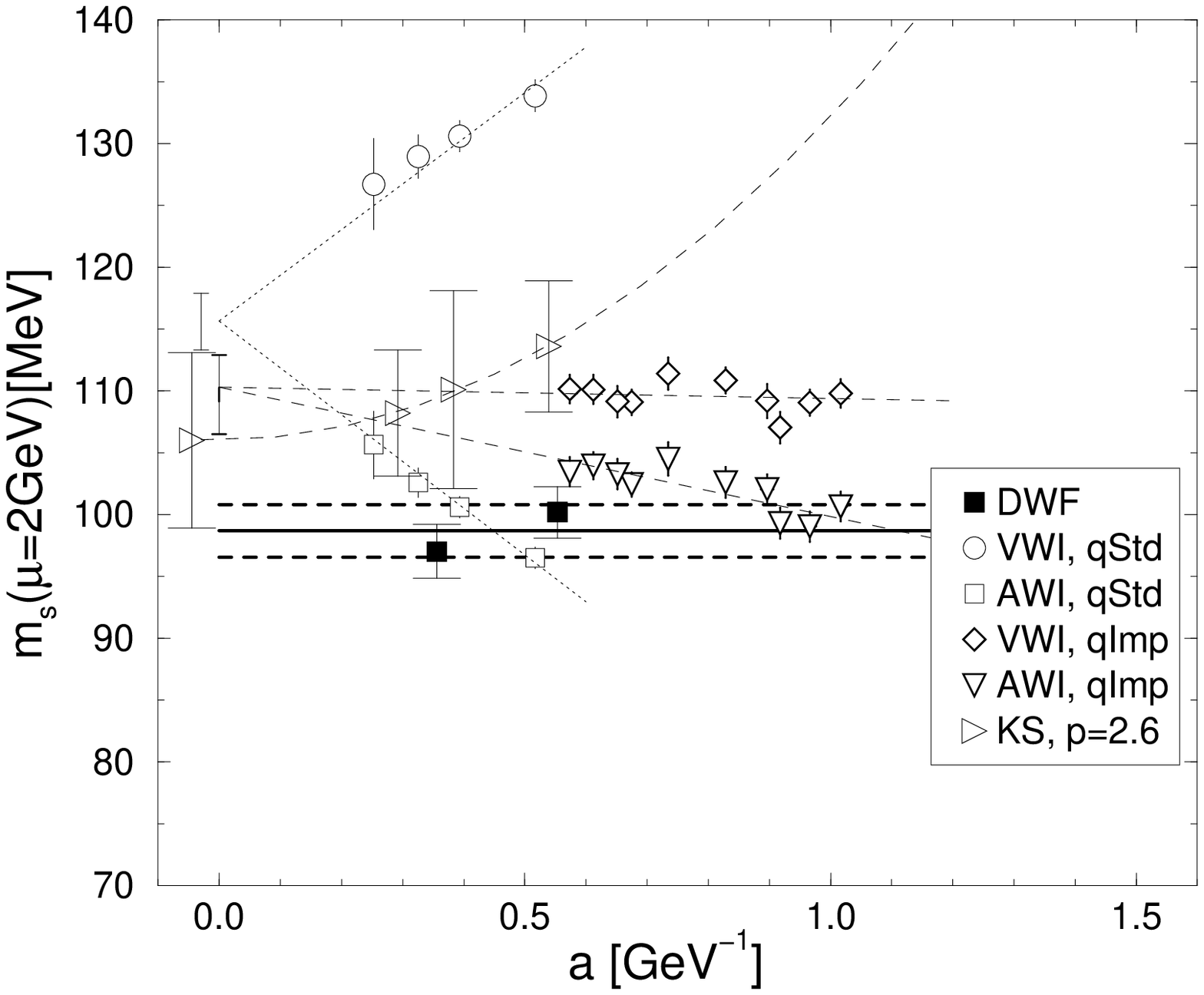}
\caption{
Scaling behavior of the strange quark mass with $K$ input, calculated
  with $m_{\rm res}$ added, compared with those from 4-dimensional quark
  action:
Wilson action (Std) \protect\cite{cppacs-quenched},
clover-improved action (Imp)\protect\cite{cppacs-full},
and Kogut-Susskind (KS) \protect\cite{jlqcd-ksqm} actions.
VWI and AWI represent vector and axial-vector Ward-Takahashi identity masses,
respectively.
$p$ for the KS fermion represents the matching scale of the RI scheme in 
  units of GeV.}
\label{fig:ms-a}
 \end{center}
\end{figure}

\begin{figure}[tb]
 \begin{center}
  \leavevmode
  \epsfxsize=9cm \epsfbox{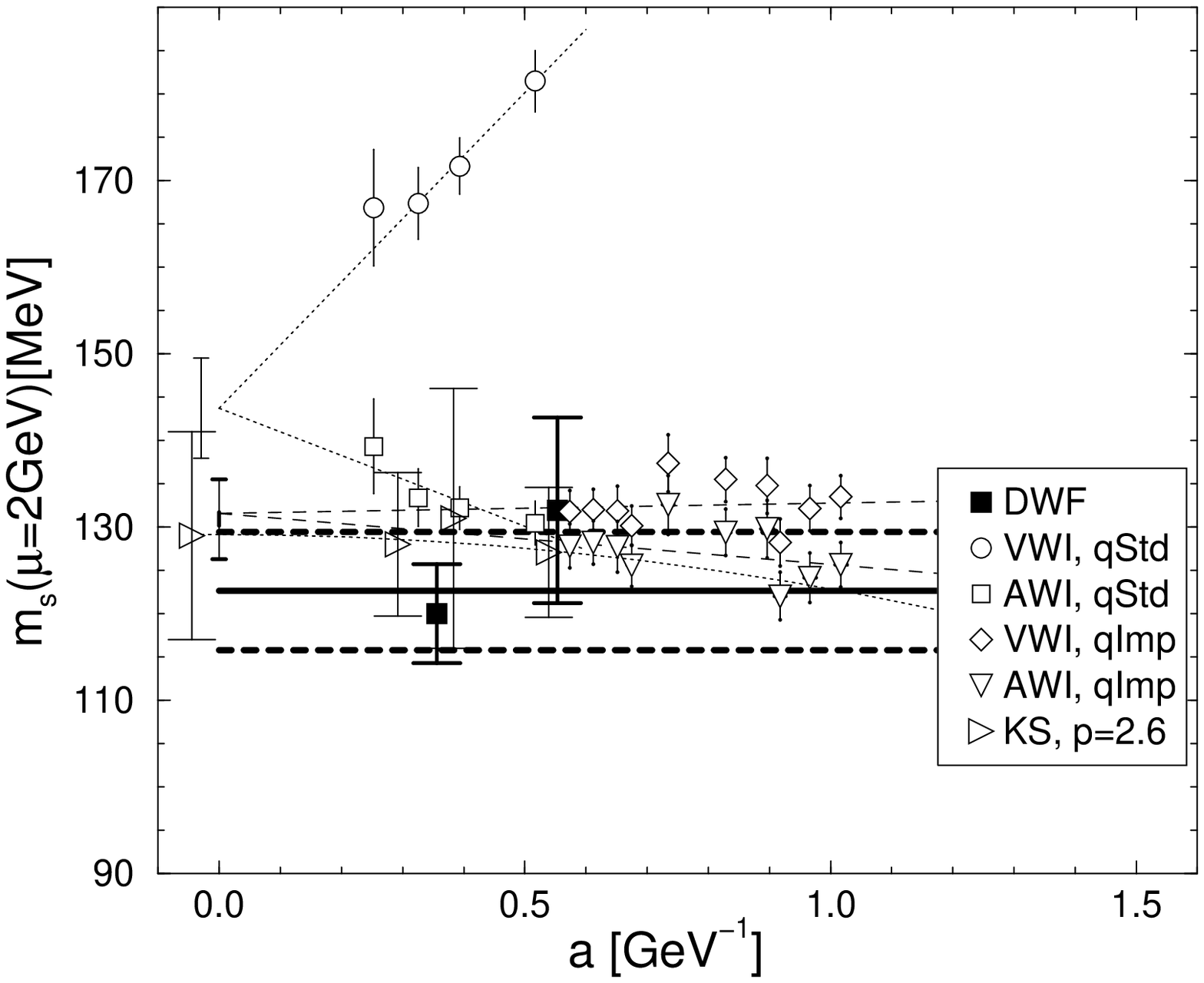}
\caption{
Scaling behavior of the strange quark mass with $\phi$ input,
 compared with those from 4-dimensional quark action:
Wilson action (Std) \protect\cite{cppacs-quenched},
clover-improved action (Imp)\protect\cite{cppacs-full},
and Kogut-Susskind (KS) \protect\cite{jlqcd-ksqm} actions.
VWI and AWI represent vector and axial-vector Ward-Takahashi identity masses,
respectively.
$p$ for the KS fermion represents the matching scale of the RI scheme in 
  units of GeV.}
\label{fig:ms-a_phi}
 \end{center}
\end{figure}

For the heavier $s$ quark, effects of $m_{res}$ are less significant, 
and those of $N_5$ are within the statistical error, as one can see in 
Table \ref{tab:qmresults}.  
In parallel with $u$-$d$ quark mass we take the results including $m_{res}$ 
as our best estimate.  The two values from our main runs at $\beta=2.6$ and 
2.9 are plotted by filled squares in Fig.~\ref{fig:ms-a} ($K$ input) and 
\ref{fig:ms-a_phi} ($\phi$ input). 
Fitting with a constant and making estimation of 
systematic errors as for $m_{ud}$ we obtain
\begin{equation}
m_s^{\overline{\rm MS}}(2 {\rm MeV})=\cases{
98.7(2.1)(5.6) {\rm MeV}  \quad K\ {\rm input}\cr
122.6(6.8)(13) {\rm MeV} \quad \phi\ {\rm input}}.
\end{equation}

In Figs.~\ref{fig:ml-a}, \ref{fig:ms-a} and \ref{fig:ms-a_phi}
open symbols show results 
obtained with the conventional 4-dimensional quark actions; 
circles and squares for the Wilson action with the plaquette gluon action 
\cite{cppacs-quenched}, diamonds and down triangles for the clover action
with the RG-improved gluon action as used in the present work
\cite{cppacs-full},
and right triangles for the Kogut-Susskind quark action with the 
plaquette gluon action \cite{jlqcd-ksqm}.
The first two cases use one-loop renormalization 
factors, while the Kogut-Susskind results are based on a non-perturbative 
value calculated in the RI scheme.  
The values estimated in the continuum limit in each of these studies 
are plotted at $a=0$ and are summarized in Table~\ref{tab:qmsummary}.  

Compared to the values obtained with the 4-dimensional quark actions, 
our results with the domain-wall action are somewhat small both for 
$u$-$d$ and $s$ quark. As with the case of $B_K$, a more precise 
examination of the issue of agreement of the continuum value requires 
a non-perturbative determination of the renormalization factors for 
our combination of quark and gluon actions.  

We note that a recent result $m_s =$ 110(2)(22) MeV \cite{RikenBNL} 
with $K$ input 
using domain wall fermions and non-perturbative renormalization 
factor but with the plaquette gauge action at $\beta=$ 6.0 is consistent 
with ours, within the 20\% systematic error quoted which includes that 
associated with the conversion from the RI scheme to the $\msbar$ scheme.

\section{Conclusions}
\label{sec:conclusion}

In this article we have presented our investigation of quenched 
calculation of the kaon B parameter $B_K$ with domain-wall QCD.
 
In order to make full use of the good chiral property of this system, 
we employed a renormalization-group improved gluon action and carried out 
simulations at $a^{-1}\gsim 2$~GeV.   According to our previous 
study\cite{cppacs-dwf}, the magnitude of chiral symmetry breaking 
due to finite fifth dimensional size $N_5$, if measured in terms of 
residual quark mass $m_{5q}$, is less than 1~MeV for $N_5\gsim 10$ at such 
lattice spacings. An explicit examination of the $N_5$ dependence of $B_K$ 
has shown that such effect is less than 1\% for $N_5\geq 16$ at 
$a^{-1}\gsim 2$~GeV.  

We have also found that spatial size effects are less than 1\% for 
the physical spatial sizes $aN_\sigma\gsim 2.5$~fm, confirming the finding 
of a previous study with the Kogut-Susskind quark action. 
Furthermore, scaling violation turned out to be very small, being less than 
2\% between $a^{-1}\approx 2$~GeV and 3~GeV.  

These results show that DWQCD, albeit computer time consuming by a factor 
${\cal O}(N_5)$ compared to conventional lattice QCD simulations, 
provides a very good framework for a precision determination of $B_K$. 
An important ingredient toward this goal,  
which was not available for the present study, 
is the value of the renormalization factors precise to the level of one
percent.
Results using the RI scheme have been reported for the plaquette action 
by the RBC Collaboration\cite{RBC}, 
and an attempt employing the Schr\"odinger functional technique is in 
progress \cite{Saoki}.  Hopefully progress in these calculations will
allow us to report results for $B_K$ in the continuum limit with a total
error of at most a few percent in quenched QCD in the near future. 

We have also examined the possibility of calculating light quark masses 
in DWQCD.  We find good scaling behavior, and the values estimated 
for the continuum limit are in reasonable agreement, albeit somewhat small, 
with those of 4-dimensional simulations.  Further progress toward 
precision determination of light quark masses also requires that 
of renormalization factors at the few percent level.

\section*{Acknowledgments}

This work is supported in part by Grants-in-Aid
of the Ministry of Education (Nos.
10640246, 10640248, 
10740107, 
11640250, 11640294, 
11740162,
12014202, 12304011, 12640253, 12740133, 13640260
).
SE, TK, KN, JN
and HPS are JSPS Research Fellows.
AAK is supported by JSPS Research for the Future Program
(No. JSPS-RFTF 97P01102).

\newcommand{\J}[4]{{#1} {#2} #3 (#4)}
\newcommand{\AP}{Ann.~Phys.}
\newcommand{\CMP}{Commun.~Math.~Phys.}
\newcommand{\IJMP}{Int.~J.~Mod.~Phys.}
\newcommand{\MPL}{Mod.~Phys.~Lett.}
\newcommand{\NP}{Nucl.~Phys.}
\newcommand{\NPSup}{Nucl.~Phys.~B (Proc.~Suppl.)}
\newcommand{\PL}{Phys.~Lett.}
\newcommand{\PR}{Phys.~Rev.}
\newcommand{\PRL}{Phys.~Rev.~Lett.}
\newcommand{\PTP}{Prog. Theor. Phys.}
\newcommand{\Suppl}{Prog. Theor. Phys. Suppl.}

\begin{table}[tb]
  \leavevmode
\begin{center}
\begin{tabular}{c|cccc|cc}
\hline
$\beta$  & 2.6 & 2.6 & 2.6 & 2.6 & 2.9 & 2.9 \\
\hline
$a^{-1}$(GeV)  
 & $   1.875(56)$ & $   1.807(37)$ & $   1.758(51)$ & $   1.847(43)$
 & $   2.869(68)$ & $   2.807(55)$ \\
\hline
$N_t$      & 40 & 40 & 40 & 40 & 60 & 60 \\
$N_\sigma$ & 16 & 24 & 24 & 32 & 24 & 32 \\
$N_5$      & 16 & 16 & 32 & 16 & 16 & 16 \\
\hline
$N_\sigma a$ (fm) & 1.7 & 2.6 & 2.6 & 3.4 & 1.7 & 2.3 \\
\hline
\# conf. &{\bf122} & {\bf 76} & {\bf 50} & {\bf 25} & {\bf 76} & {\bf 50} \\
\hline
\end{tabular}
\caption{
Simulation parameters together with the number of configurations 
analyzed shown in bold letters. 
}
\label{tab:config}
\end{center}
\end{table}

\begin{table}[p]
  \leavevmode
\begin{center}
\begin{tabular}{l|cccccc}
 $M$
 & $z_2$ & $z_m$ & $z_w$ & $z_{A}$ & $z_{P}$ & $z_{O_4}$ \cr
\hline
$1.8$
 & $-3.824$ & $13.148$ & $-25.1295$ & $-9.190$ & $-13.148$ & $-23.868$\cr
\hline
\hline
 $M$
 & $z_2^{\rm MF}$ & $z_m^{\rm MF}$ & $z_w^{\rm MF}$
 & $z_{A}^{\rm MF}$ & $z_{P}^{\rm MF}$ & $z_{O_4}^{\rm MF}$ \cr
\hline
$1.4198$
 & $0.651$ & $6.044$ & $-7.92355$ & $-4.692$ & $-6.044$ & $-13.612$\cr
$1.4687$
 & $0.632$ & $6.319$ & $-7.95874$ & $-4.714$ & $-6.319$ & $-13.500$\cr
\end{tabular}
\caption{Finite parts of the renormalization factors with RG improved
 gauge action.
The mean field approximation is used for the factors at $M=1.4198$ and
 $1.4687$.
Errors from the numerical integration are in the last written digit.
}
\label{tab:zfactor}
\end{center}
\end{table}

\begin{table}[p]
  \leavevmode
\begin{center}
\begin{tabular}{l|c|cc}
 $M$
 & $z_{B_K}$ & $z_{B_P}$ & $z_{B_P}^{\rm MF}$ \cr
\hline
$1.8$
 & $0.64$  & $11.19$ & $-$ \cr
$1.4198$
 & $-1.10$ & $-$ & $2.51$ \cr
$1.4687$
 & $-0.93$ & $-$ & $3.35$ \cr
\end{tabular}
\caption{Finite parts of the renormalization factors $z_{B_K}$ and
 $z_{B_P}$ with RG improved gauge action.
The mean field approximation is used for $z_{B_P}$ at $M=1.4198$ and
 $1.4687$, while the effect of the approximation on $z_{B_K}$ is just to 
 shift the domain-wall height.
Errors from the numerical integration are in the last written digit.}
\label{tab:zbk}
\end{center}
\end{table}

\begin{table}[p]
  \leavevmode
\begin{center}
\begin{tabular}{l|ccccc}
 & \multicolumn{4}{c}{$\beta=2.6$ on $16^3\times40\times16$ lattice} \cr
\hline
 $m_f$ & $B_K$ & $B_P$ & $m_{PS}$ & $m_V$ & $m_{PS}/m_V$\cr
\hline
$0.01$ & $   0.487(19)$ & $  0.0225(10)$ & $  0.1902(29)$ & $   0.438(12)$ &
 $   0.434(13)$ \cr
$0.02$ & $   0.566(14)$ & $  0.0569(14)$ & $  0.2554(23)$ & $  0.4639(72)$ &
 $  0.5505(97)$ \cr
$0.03$ & $   0.615(11)$ & $  0.0945(17)$ & $  0.3083(21)$ & $  0.4904(54)$ &
 $  0.6287(80)$ \cr
$0.04$ & $  0.6486(85)$ & $  0.1327(19)$ & $  0.3548(20)$ & $  0.5168(43)$ &
 $  0.6865(69)$ \cr
\end{tabular}
\caption{Data for $B_K$, $B_P$, $m_{PS}$, $m_V$ and $m_{PS}/m_V$
 at each quark mass $m_f$ at $\beta=2.6$ on $16^3\times40\times16$ lattice.}
\label{tab:R26.16x40x16}
\end{center}
\end{table}
\begin{table}[p]
  \leavevmode
\begin{center}
\begin{tabular}{l|ccccc}
 & \multicolumn{4}{c}{$\beta=2.6$ on $24^3\times40\times16$ lattice} \cr
\hline
 $m_f$ & $B_K$ & $B_P$ & $m_{PS}$ & $m_V$ & $m_{PS}/m_V$\cr
\hline
$0.01$ & $  0.5196(69)$ & $ 0.02487(38)$ & $  0.1883(13)$ & $  0.4522(83)$ &
 $  0.4164(80)$ \cr
$0.02$ & $  0.5769(47)$ & $ 0.05777(55)$ & $  0.2567(12)$ & $  0.4683(51)$ &
 $  0.5481(65)$ \cr
$0.03$ & $  0.6155(42)$ & $ 0.09367(70)$ & $  0.3102(11)$ & $  0.4900(36)$ &
 $  0.6331(52)$ \cr
$0.04$ & $  0.6453(39)$ & $ 0.13092(83)$ & $ 0.35647(97)$ & $  0.5141(29)$ &
 $  0.6934(42)$ \cr
\end{tabular}
\caption{Data for $B_K$, $B_P$, $m_{PS}$, $m_V$ and $m_{PS}/m_V$
 at each quark mass $m_f$ at $\beta=2.6$ on $24^3\times40\times16$ lattice.}
\label{tab:R26.24x40x16}
\end{center}
\end{table}
\begin{table}[p]
  \leavevmode
\begin{center}
\begin{tabular}{l|ccccc}
 & \multicolumn{4}{c}{$\beta=2.6$ on $24^3\times40\times32$ lattice} \cr
\hline
 $m_f$ & $B_K$ & $B_P$ & $m_{PS}$ & $m_V$ & $m_{PS}/m_V$\cr
\hline
$0.01$ & $  0.5225(89)$ & $ 0.02444(47)$ & $  0.1874(16)$ & $   0.463(12)$ &
 $   0.405(11)$ \cr
$0.02$ & $  0.5780(63)$ & $ 0.05751(71)$ & $  0.2558(14)$ & $  0.4760(73)$ &
 $  0.5374(89)$ \cr
$0.03$ & $  0.6152(54)$ & $ 0.09329(90)$ & $  0.3093(12)$ & $  0.4956(50)$ &
 $  0.6242(67)$ \cr
$0.04$ & $  0.6437(49)$ & $  0.1302(10)$ & $  0.3556(11)$ & $  0.5182(37)$ &
 $  0.6862(53)$ \cr
\end{tabular}
\caption{Data for $B_K$, $B_P$, $m_{PS}$, $m_V$ and $m_{PS}/m_V$
 at each quark mass $m_f$ at $\beta=2.6$ on $24^3\times40\times32$ lattice.}
\label{tab:R26.24x40x32}
\end{center}
\end{table}
\begin{table}[p]
  \leavevmode
\begin{center}
\begin{tabular}{l|ccccc}
 & \multicolumn{4}{c}{$\beta=2.6$ on $32^3\times40\times16$ lattice} \cr
\hline
 $m_f$ & $B_K$ & $B_P$ & $m_{PS}$ & $m_V$ & $m_{PS}/m_V$\cr
\hline
$0.01$ & $  0.5260(47)$ & $ 0.02594(33)$ & $  0.1841(13)$ & $  0.4424(92)$ &
 $  0.4161(86)$ \cr
$0.02$ & $  0.5766(40)$ & $ 0.05873(51)$ & $  0.2535(12)$ & $  0.4608(59)$ &
 $  0.5501(77)$ \cr
$0.03$ & $  0.6124(39)$ & $ 0.09460(72)$ & $  0.3076(12)$ & $  0.4847(44)$ &
 $  0.6347(66)$ \cr
$0.04$ & $  0.6424(35)$ & $ 0.13196(88)$ & $  0.3544(11)$ & $  0.5097(36)$ &
 $  0.6954(57)$ \cr
\end{tabular}
\caption{Data for $B_K$, $B_P$, $m_{PS}$, $m_V$ and $m_{PS}/m_V$
 at each quark mass $m_f$ at $\beta=2.6$ on $32^3\times40\times16$ lattice.}
\label{tab:R26.32x40x16}
\end{center}
\end{table}
\begin{table}[p]
  \leavevmode
\begin{center}
\begin{tabular}{l|ccccc}
 & \multicolumn{4}{c}{$\beta=2.9$ on $24^3\times60\times16$ lattice} \cr
\hline
 $m_f$ & $B_K$ & $B_P$ & $m_{PS}$ & $m_V$ & $m_{PS}/m_V$\cr
\hline
$0.01$ & $   0.512(18)$ & $  0.0340(14)$ & $  0.1444(20)$ & $  0.2915(59)$ &
 $   0.496(13)$ \cr
$0.02$ & $   0.591(11)$ & $  0.0802(17)$ & $  0.2007(17)$ & $  0.3172(40)$ &
 $  0.6328(96)$ \cr
$0.03$ & $  0.6403(84)$ & $  0.1295(21)$ & $  0.2460(15)$ & $  0.3425(32)$ &
 $  0.7182(78)$ \cr
$0.04$ & $  0.6758(75)$ & $  0.1793(24)$ & $  0.2856(14)$ & $  0.3681(27)$ &
 $  0.7760(65)$ \cr
\end{tabular}
\caption{Data for $B_K$, $B_P$, $m_{PS}$, $m_V$ and $m_{PS}/m_V$
 at each quark mass $m_f$ at $\beta=2.9$ on $24^3\times60\times16$ lattice.}
\label{tab:R29.24x60x16}
\end{center}
\end{table}
\begin{table}[p]
  \leavevmode
\begin{center}
\begin{tabular}{l|ccccc}
 & \multicolumn{4}{c}{$\beta=2.9$ on $32^3\times60\times16$ lattice} \cr
\hline
 $m_f$ & $B_K$ & $B_P$ & $m_{PS}$ & $m_V$ & $m_{PS}/m_V$\cr
\hline
$0.01$ & $  0.5318(72)$ & $ 0.03412(64)$ & $  0.1459(12)$ & $  0.2984(54)$ &
 $   0.489(10)$ \cr
$0.02$ & $  0.5922(58)$ & $ 0.07885(97)$ & $  0.2022(12)$ & $  0.3209(31)$ &
 $  0.6302(78)$ \cr
$0.03$ & $  0.6345(51)$ & $  0.1271(12)$ & $  0.2470(12)$ & $  0.3451(23)$ &
 $  0.7157(65)$ \cr
$0.04$ & $  0.6669(46)$ & $  0.1761(14)$ & $  0.2863(11)$ & $  0.3700(19)$ &
 $  0.7738(55)$ \cr
\end{tabular}
\caption{Data for $B_K$, $B_P$, $m_{PS}$, $m_V$ and $m_{PS}/m_V$
 at each quark mass $m_f$ at $\beta=2.9$ on $32^3\times60\times16$ lattice.}
\label{tab:R29.32x60x16}
\end{center}
\end{table}

\begin{table}[p]
  \leavevmode
\begin{center}
\begin{tabular}{c|cccc|cc}
\hline
$\beta$  & 2.6 & 2.6 & 2.6 & 2.6 & 2.9 & 2.9 \\
\hline
$N_t$      & 40 & 40 & 40 & 40 & 60 & 60 \\
$N_\sigma$ & 16 & 24 & 24 & 32 & 24 & 32 \\
$N_5$      & 16 & 16 & 32 & 16 & 16 & 16 \\
\hline
$a^{-1}$(GeV)
 & $   1.875(56)$ & $   1.807(37)$
 & $   1.758(51)$ & $   1.847(43)$
 & $   2.869(68)$ & $   2.807(55)$
 \cr
\hline
$m_V(m_f=0)a$
 & $   0.411(12)$ & $  0.4261(87)$
 & $   0.438(13)$ & $  0.4169(97)$
 & $  0.2684(64)$ & $  0.2743(54)$ \cr
$m_{PS}^2(m_f=0)a^2$
 & $  0.0060(12)$ & $ 0.00490(53)$
 & $ 0.00467(69)$ & $ 0.00330(52)$
 & $ 0.00038(62)$ & $ 0.00099(34)$ \cr
\hline
$m_{\rm res}a$
 & $ 0.00201(42)$ & $ 0.00161(18)$
 & $ 0.00153(23)$ & $ 0.00108(17)$
 & $ 0.00019(31)$ & $ 0.00049(17)$
 \cr
$m_{\rm res}$ (MeV)
 & $    3.77(78)$ & $    2.90(32)$
 & $    2.70(42)$ & $    1.99(32)$
 & $    0.54(89)$ & $    1.38(48)$
 \cr
$m_f^{ud}a$
 & $-0.00027(41)$ & $ 0.00022(19)$
 & $ 0.00040(26)$ & $ 0.00067(18)$
 & $ 0.00091(32)$ & $ 0.00066(18)$
 \cr
$m_f^sa/2 (K)$
 & $  0.0216(15)$ & $  0.0233(10)$
 & $  0.0248(15)$ & $  0.0227(11)$
 & $ 0.01475(82)$ & $ 0.01515(70)$
 \cr
$m_f^sa (\phi)$
 & $  0.0503(62)$ & $  0.0632(64)$
 & $   0.071(11)$ & $  0.0583(59)$
 & $  0.0350(25)$ & $  0.0373(25)$ \cr
\end{tabular}
\caption{Results of meson mass fits.}  
\label{tab:massresults}
\end{center}
\end{table}

\begin{table}[p]
  \leavevmode
\begin{center}
\begin{tabular}{c|cccc|cc}
\hline
$\beta$  & 2.6 & 2.6 & 2.6 & 2.6 & 2.9 & 2.9 \\
\hline
$N_t$      & 40 & 40 & 40 & 40 & 60 & 60 \\
$N_\sigma$ & 16 & 24 & 24 & 32 & 24 & 32 \\
$N_5$      & 16 & 16 & 32 & 16 & 16 & 16 \\
\hline
$a^{-1}$(GeV)
 & $   1.875(56)$ & $   1.807(37)$ & $   1.758(51)$ & $   1.847(43)$
 & $   2.869(68)$ & $   2.807(55)$ \\
\hline
\multicolumn{7}{l}{bare $B$ parameters}\\
$B_K$
 & $   0.575(14)$ & $  0.5908(57)$ & $  0.5975(77)$ & $  0.5871(60)$
 & $   0.554(14)$ & $  0.5655(69)$ \cr
\hline
\multicolumn{7}{l}{renormalized $B$ parameters 
($\overline{\rm MS}$ scheme with NDR at 
$\mu=2$~GeV)}\\
$B_K(q^*=1/a)$
& $   0.564(14)$ & $  0.5782(55)$
& $  0.5839(75)$ & $  0.5753(58)$
& $   0.558(14)$ & $  0.5690(70)$ \cr
$B_K(q^*=\pi/a)$
& $   0.570(14)$ & $  0.5844(56)$
& $  0.5901(76)$ & $  0.5815(59)$
& $   0.563(15)$ & $  0.5741(70)$ \cr
$B_K(g=g_P)$
& $   0.566(14)$ & $  0.5803(56)$
& $  0.5862(75)$ & $  0.5773(59)$
& $   0.557(14)$ & $  0.5684(70)$ \cr
\end{tabular}
\caption{Results for $B$ parameters together with those of relevant 
quantities. $g=g_P$ in the last row denotes the use of an alternative 
definition of the coupling (\protect\ref{eqn:coupling-P}).}  

\label{tab:results}
\end{center}
\end{table}

\begin{table}[p]
  \leavevmode
\begin{center}

\begin{tabular}{c|cc|c}
parameters  & $B_K$  & error  & $B_K$ \\
\hline
$B$ & & & 0.41180 \\
$b$ & & & 0.71110 \\
$c$ & & & 0.20731 \\
\hline
$m_{PS}^2$ (GeV$^2$) &\multicolumn{2}{c|}{continuum extrapolation} & 
reconstruction by the fit \\
\hline
     0.020 &     0.4318 &     0.0100 &     0.4377  \\
     0.100 &     0.4994 &     0.0058 &     0.5001  \\
     0.200 &     0.5544 &     0.0044 &     0.5528  \\
     0.300 &     0.5937 &     0.0043 &     0.5922  \\
     0.400 &     0.6239 &     0.0041 &     0.6228  \\
     0.500 &     0.6471 &     0.0038 &     0.6470  \\
     0.600 &     0.6643 &     0.0038 &     0.6660  \\
     0.700 &     0.6780 &     0.0043 &     0.6807  \\
     0.800 &     0.6907 &     0.0051 &     0.6918  \\
     0.900 &     0.7016 &     0.0060 &     0.6996  \\
     1.000 &     0.7097 &     0.0075 &     0.7046  \\
\end{tabular}
\caption{Parameters for the fit of $B_K$ 
in the continuum limit by Eq.~(\protect\ref{eq:bkmpi}), and
$B_K$ as a function of $m_{PS}^2$ in the continuum limit,
together with the reconstruction from the fit.}

\label{tab:contbk}
\end{center}
\end{table}
\begin{table}[p]
  \leavevmode
\begin{center}
\begin{tabular}{c|cccc|cc}
$\beta$  & 2.6 & 2.6 & 2.6 & 2.6 & 2.9 & 2.9 \\
\hline
$N_t$      & 40 & 40 & 40 & 40 & 60 & 60 \\
$N_\sigma$ & 16 & 24 & 24 & 32 & 24 & 32 \\
$N_5$      & 16 & 16 & 32 & 16 & 16 & 16 \\
\hline
$a^{-1}$(GeV)
& $   1.875(56)$ & $   1.807(37)$
& $   1.758(51)$ & $   1.847(43)$
& $   2.869(68)$ & $   2.807(55)$ \cr
\hline
\multicolumn{7}{l}{renormalized quark masses ignoring $m_{\rm res}$}\\
$m_{u,d}$(MeV)
& $   -0.60(90)$ & $    0.47(39)$
& $    0.82(52)$ & $    1.44(37)$
& $     3.0(10)$ & $    2.12(57)$ \cr
$m_s$(MeV)
& $    94.9(33)$ & $    96.8(22)$
& $    99.8(31)$ & $    95.9(24)$
& $    94.4(27)$ & $    95.4(23)$ \cr
\hline
\multicolumn{7}{l}{renormalized quark masses including $m_{\rm res}$}\\
$m_{u,d}+m_{\rm res}$(MeV)
& $    3.79(12)$ & $   3.821(79)$
& $    3.92(11)$ & $   3.748(92)$
& $   3.625(93)$ & $   3.701(83)$ \cr
$m_s+m_{\rm res}$(MeV)
& $    99.3(33)$ & $   100.2(21)$
& $   102.9(30)$ & $    98.3(24)$
& $    95.0(24)$ & $    97.0(22)$ \cr
\hline
\multicolumn{7}{l}{renormalized $s$ quark mass with $\phi$ input}\\
$m_s$(MeV)
& $    110.(10)$ & $    132.(11)$
& $    144.(19)$ & $   124.8(98)$
& $   115.6(58)$ & $   120.0(57)$ \cr
\end{tabular}
\caption{Results for quark masses.  Renormalized values are in 
the $\overline{\rm MS}$ scheme at $\mu=2$~GeV.}
\label{tab:qmresults}
\end{center}
\end{table}

\begin{table}[p]
  \leavevmode
\begin{center}
\begin{tabular}{ccc|ccc}
ref. &quark action & gluon action & $m_{ud}^{\overline{\rm MS}}(2 {\rm GeV})$ &
\multicolumn{2}{c}{$m_{s}^{\overline{\rm MS}}(2 {\rm GeV})$} \\
&&&& $K$ input & $\phi$ input\\
\hline
this work & DW & RG-improved & 3.764(81)(215) MeV & 98.7(2.1)(5.6) MeV 
                                                & 122.6(6.8)(13) MeV \\
\hline
\protect\cite{cppacs-quenched}&
Wilson  & plaquette   & 4.57(18) MeV      & 116 (3) MeV & 144 (6) MeV \\
\protect\cite{cppacs-full}&
clover  & RG-improved & $4.36^{+0.14}_{-0.17}$ MeV &
                             $110^{+3}_{-4}$ MeV & $132^{+4}_{-6}$ MeV \\
\protect\cite{jlqcd-ksqm}&
KS      & plaquette   & 4.23(29) MeV      & 106 (7) MeV  & 129 (12) MeV \\
\hline
\end{tabular}
\caption{Results for light quark masses as compared with 
previous studies.  One-loop approximation to the renormalization factors 
are employed except for those with the KS fermion action in the last row. }
\label{tab:qmsummary}
\end{center}
\end{table}

\end{document}